\documentclass{aa}
\usepackage{graphicx}
\usepackage{amssymb}
\usepackage{natbib}

\begin{document}
   \title{Stellar evolution with rotation XIII: }

\subtitle{Predicted GRB rates at various Z.}

   \author{R. Hirschi \inst{1}
          \and
          G. Meynet \inst{2}
          \and
          A. Maeder \inst{2}
          }

   \offprints{raphael.hirschi@unibas.ch}

   \institute{Dept. of physics and Astronomy, University of Basel,           
	      Klingelbergstr. 82, CH-4056 Basel
              \and
     	      Geneva Observatory, CH--1290 Sauverny, Switzerland\\
             }


   \date{Received / Accepted }

\abstract{

We present the evolution of rotation in models of massive single
stars covering
a wide range of masses and metallicities.
These models reproduce very well observations during the early stages of
the evolution \citep[in particular WR populations and ratio between type II and type Ib,c at
different metallicities, see][]{ROTXI}.  

Our models predict the production of fast rotating black holes. 
Models with large initial masses or high metallicity end their life 
with less angular momentum
in their central remnant with respect to the break--up limit for the remnant.
Many WR star models satisfy the three main criteria (black hole formation,
loss of hydrogen--rich envelope and enough angular momentum to form an
accretion disk around the black hole) for gamma--ray bursts (GRB) 
production via the collapsar
model \citep{W93}. Considering all types of WR stars as GRB progenitors, 
there would be too many GRBs compared to observations. If we consider only 
WO stars \citep[type
Ic supernovae as is the case for SN2003dh/GRB030329, see][]{MMN03} as
GRBs progenitors, the GRBs production rates are in much better agreement
with observations. WO stars are produced only at low metallicities in the
present grid of models. This prediction can be tested by future
observations. 
}

   \maketitle

\section{Introduction}
In the previous papers in the series on stellar evolution with rotation,
the effects of rotation on the stellar evolution was studied with an 
emphasis on the early stages of the pre--supernova evolution (main
sequence, MS, and He--burning). The
models reproduce very well many observational features at various 
metallicities, like
surface enrichments \citep{MM02n}, ratios between red and blue 
supergiants \citep{ROTVII} (hereinafter paper VII).
In \citet{ROTXI} (paper XI) and 
\citet{ROTX} (paper X), the Wolf--Rayet (WR hereinafter) 
star population at
different metallicities are studied. 
In view of the lower mass loss rates obtained when clumping effects in
the winds are accounted for, models without rotation do not reproduce
the WR populations in galaxies.
Models of rotating massive stars
give much better fit to the populations of WR stars at different
metallicities than non--rotating models. They can reproduce the 
observed number ratio of WR to 
O-type stars, the observed ratio of WN to WC stars (for metallicities
lower than solar), the observed fraction of WR stars in the transition
phase WN/WC and finally the observed ratio of type Ib and Ic to type II
supernovae. A good fit of the observed properties of WR stars is of
particular interest in this study since WR stars are thought to be 
progenitors of GRBs.

In \citet{psn04a} (hereinafter paper XII), we described 
the recent modifications made to the Geneva code
and the evolution of the models until silicon burning.
In this paper, we look at the evolution of rotation in massive stars
with an emphasis on the final stages of the evolution towards 
neutron stars (NS), black holes (BH) and especially 
long soft $\gamma-$ray bursts (GRB).

\subsection{Neutrons stars}
There are many observations of neutron stars and pulsars available. 
A catalogue of observed pulsars and their properties is available at the
web page of the group \citet{PSR}. The fastest young pulsars have periods
larger than 10 ms: 16 ms for J0537-6910 \citep{MG98},
33 ms for the Crab pulsar \citep[B0531+21, ][]{SR68},
50 ms for J0540-6919 \citep{SHH84}. There are about 20 pulsars with a
period smaller than 100 ms in the catalogue with an age less than 100'000
years.
The pulsars may have slightly slowed down since their birth but the
initial period of pulsars is at least 10 ms.
NS are supposed to form during the collapse of stars with an 
initial mass between about 10 and 25 $M_\odot$. 
If we compare 10 ms with the values obtained in Tables
\ref{moma04}--\ref{moma40} (columns 11 and 12), we see  
that in general our models have much
more angular momentum than the observed pulsars. The difference can
reach a factor of about 100 around 15 $M_\odot$ and a factor of about 10
around 60 $M_\odot$. 
In order to reconcile the model predictions with the observations,
additional angular momentum has to be lost before the formation of the
pulsar. This can occur during the pre--supernova stages due to the
effects of the magnetic fields not taken into account in this work or
during the collapse and the explosion.
The existence of a primordial magnetic field in some stars \citep{BS04} 
may also slow down their core. 
Models including the effects of magnetic fields have only recently been
developed \citep{HWLS03,ROTMII} due to the complexity of the interplay 
between rotation and magnetic fields.
\citet{HWS05} show that the braking by magnetic fields \citep[which are
produced by differential rotation during the evolution, see][]{Sp02} 
reduce significantly the discrepancy between the predicted and observed 
pulsar periods.
However, these models as well as binary models presented in \citet{PLYH05} contain too
little angular momentum at the pre--supernova stage to produce GRBs.

In this paper, 
we study the case
without magnetic field, which is more favorable to the GRB production.
The case with magnetic braking will be treated in a further study.
In addition, the present study offers a basis of comparison for future
models including magnetic fields.
It is important here to note that internal magnetic braking 
is most efficient during the Red SuperGiant (RSG) phase during which the star has a fast
rotating core and a slow rotating envelope.
This means that magnetic braking is much less effective for 
stars which skip the RSG stage, as mentioned in \citet{HWS05}. The stars
which skip the RSG stage are the most massive WR stars, with masses larger than
about 60 $M_\odot$. Furthermore, more massive stars have shorter
lifetimes and thus less time to be slowed down. It is also possible that
mechanisms that slow down the nascent neutron star may not be as
efficient when a black hole is formed. These differences between
stars with masses around 15 $M_\odot$ (forming a neutron star and going through the RSG stage) 
and 60 $M_\odot$ (not going through the RSG phase and forming black
holes) could possibly explain, with the same physics, the pulsar periods
and the GRBs. Further studies will verify this possibility. 

Braking between the collapse and the pulsar formation can also occur
via different mechanisms: r-Mode instability, neutrino--powered magnetic
stellar winds, fall--back and the propeller mechanism. However, these mechanisms may not be efficient
enough to slow down the core efficiently \citep[see][ for a
discussion]{HWS05}. Latest models studying gravitational waves seem to indicate that
braking is possible during and after the collapse \citep{OTB05}. 
If the cores are not slowed down during the pre--supernova stages, 
rotation may play an important role in supernova explosions.
Rotation may in this case provide a large amount of energy for the 
explosion ($\sim 10^{52}$ erg) but this is generally not observed \citep{J05}. 

\subsection{Black holes and GRBs}
Theoretical models still struggle to reproduce supernova explosions
 \citep[see][ for a discussion and
references]{FH05}. It is therefore not possible for us to predict with 
certainty or precision
the fate of the remnant of our models. Nevertheless, following the estimates 
presented in
\citet{F99} and \citet{HFWLH03}, we will consider in this study that the 
lower mass limit for black
hole formation is around 25 $M_\odot$. The upper mass limit depends
strongly on the mass loss and is around 100 $M_\odot$ at solar metallicity.
The maximum neutron star mass is also still uncertain
\citep[see][ for recent studies]{Sr02,MBS04}. 
Depending on the nuclear equation of state used and rotation rate, 
the upper mass limit for neutron stars can vary between 1.6 and 3
$M_\odot$ although 2 to 2.5 $M_\odot$ is a more common range.
Black holes cannot be observed directly. Thus they are usually detected in
X--ray binaries. In certain cases, 
observations allow the determination of the masses of
the binary companions. If the accreting object has a derived 
mass larger than 2--3 $M_\odot$, it is considered as a BH candidate. 
The best known candidates are LMC X--3 and Cyg
X--1 \citep[see][ and references therein]{KM05}.

GRBs can be divided in 
two main types: a) short and hard and b) long and soft.
See \citet{Pi04} for a recent review on GRBs.
The long soft GRBs have recently been connected to supernova
(SN) explosions of the type Ib,c \citep[see for example][]{Ma03}.
Since then, several studies have been devoted to find which stars can
lead to GRBs. \citet{HWLS03}, \citet{HWS05} and \citet{psn04a} 
looked at massive single star at solar metallicity as progenitors. 
Note that the two models differ in the treatment of meridional
circulation, which is an advective process \citep{ROTVIII}. 
Our models account for the advection of angular momentum
during the MS phase while models of Heger and collaborators treat
meridional circulation as a diffusive process. 
\citet{ROTBIII} show
that the treatment of meridional circulation as an advective process is
crucial to model the interplay between circulation and magnetic braking.
Another difference in this article is the fact that we study 
in detail
the effects of metallicity
in the context of GRBs for models of differentially rotating stars.
This is important since metallicities lower than solar are expected to be
more favourable to the production of GRBs 
\citep[due to weaker mass loss, see][]{FW99,WH03}.
\citet{PMN04} and \citet{IRT04} considered both single and binary 
massive stars. Note that \citet{IRT04} 
assume that the loss of angular momentum is proportional to the amount
of mass lost and do not consider internal transport of angular momentum.
Their approximate treatment of the evolution of angular momentum leads
them to the conclusion that only binary stars can retain enough angular
momentum in their core to form GRBs. Even without magnetic braking,
their models predict no GRBs from single massive stars which is in
contradiction with our models and those of \citet{HLW00}. 
\citet{PLYH05}
look at single and binary systems with and without magnetic fields at
solar metallicity. They find that in the models followed, both single
and binary models without magnetic field can produce GRBs and both 
single and binary models with magnetic field considered in the study 
cannot produce GRBs. They conclude that, if the present modelling of magnetic
fields is accurate, GRBs have to be produced in some exotic channels of
binary systems.
\citet{FH05} looked at massive binary star mergers, which are believed
to be one of the best binary candidates for GRBs. 
They follow the evolution 
prior to and after the merger with the KEPLER code \citep{HLW00}
and the merger process with 3D simulations, which is very interesting
and which has a lot of potential for further applications. 
They find that, in some
cases, merged helium stars can retain more angular momentum than single
massive stars. However, they obtain these faster rotators only for models in
which they remove angular momentum artificially without removing mass
from the merged star. If more reasonable mass loss prescriptions are
used after the merger, the final angular momentum contained in the 
central remnant is
similar in merged stars and in single massive stars. 
This is due to the strong mass loss that WR stars experience, especially
the fast rotating ones. Merged stars lose a large amount of mass and some
merged systems probably form NSs instead of BHs. 
It is therefore not proven that binary stars retain more angular
momentum than single stars and the exact frequency of potential binary
star progenitors has not been compared quantitatively with the GRB rates.

Here we discuss the possibility for single massive stars to be 
progenitors studying
models covering a large range of initial masses and metallicities. These
models reproduce very well the WR star population and many of them
retain enough angular momentum to produce GRBs as shown below. We described the
calculations in Sect. 2, the evolution of rotation in Sect. 3 and present our predicted
GRB rates in Sect. 4.
\section{Description of the calculations}
In order to study the evolution of core rotation and its dependence on initial mass
and metallicity, we used the models presented in 
\citet{ROTX} (paper X) and \citet{psn04a} (paper XII) for solar metallicity
and 
\citet{ROTXI} (paper XI) for non--solar metallicities, 
which span a wide range in initial masses 
(9 to 120\,$M_\odot$ at solar metallicity) and metallicities 
($Z=0.004$ to $Z=0.040$). 

All these models (except for one 60 $M_\odot$ model at $Z=0.004$) 
have an initial rotational velocity of 300\,km\,s$^{-1}$, 
which gives an average value on the MS of about 220\,km\,s$^{-1}$, 
the average observed rotational velocities on the MS
\citep[see for example][]{FU82}. They were all calculated with the
same input physics: Schwarzschild criterion for convection, overshooting of 0.1 H$_{\rm P}$ for the
hydrogen (H) and helium (He) burning cores, same mass loss prescriptions, 
treatment of rotation
including meridional circulation and shear diffusion
(see papers X--XII for more details). 

Non-solar metallicity models have been evolved until the end of core He--burning. 
Solar metallicity models have been evolved until the end of core He--burning for the 
9, 85 and 120 $M_\odot$ models, end of oxygen (O) burning 
for the 12 $M_\odot$ model and end of
silicon (Si) burning for the 15, 20, 25, 40 and 60 $M_\odot$ models.
As we shall see later, the largest changes in the angular momentum of the core occurs
during H and He--burnings and this is why we did not evolve non-solar 
metallicity models
further than the end of He--burning.

Several quantities are useful for studying the evolution of rotation:

- The angular velocity at mass coordinate $m$ or radius $r$ \citep[here $r$ is the average
radius of a shell, see][ for details]{ROTI}, $\Omega$ [$\rm s^{-1}$], 
and its ratio to the Keplerian velocity,
$\Omega/\Omega_K$, where $\Omega_K=\sqrt{G\,m/r^3}$.
When $\Omega/\Omega_K$ becomes larger than about 0.9, 
the star gets elongated along its
equator and at break--up, the star equatorial radius is 1.5 times larger than the polar
radius (value obtained using the Roche model). 
Therefore the star may reach break--up  
before $\Omega/\Omega_K$ approaches one, in particular for models with a
large Eddington factor which reach break--up for $\Omega$ much lower 
than $\Omega_K$ \citep{ROTVI}.

- The angular momentum
of a core of mass $M$, $\cal{L}$$_M=\int_0^M j_m\,dm$ [$\rm g\,cm^2\,s^{-1}$],
where $j_m=2/3\,\Omega_m\,r_m^2$ [$\rm cm^2\,s^{-1}$] is the specific angular momentum at
mass coordinate $m$. The momentum of inertia of a core of mass $M$,
$\cal{I}$$_M=\int_0^M 2/3\,r^2\,dm$ [$\rm g\,cm^2$].

The properties of the models are presented in four Tables: 
Table \ref{moma04} for the metallicity of the Small Magellanic Cloud (SMC, Z=0.004),
Table \ref{moma08} for the Large Magellanic Cloud (LMC, Z=0.008),
Table \ref{moma20} for solar metallicity (Z=0.020) and
Table \ref{moma40} for the Galactic center (GC, Z=0.040).
In these tables, 
for each model, we give the initial mass and velocity as well as the remnant
mass estimated from the carbon--oxygen (CO) core mass, using the relation from
\citet{AM92}. We used the value of the CO core at the end of the evolution for the
calculation (at the end of core He--burning in general and at the end of
Si--burning for the models from paper XII).
Then,
\begin{itemize}
\item column 1 is the evolutionary stage to which the values correspond,
\item column 2 is the total mass at the given stage,
\item column 3 is the angular momentum contained in the remnant,
${\cal L}_{\rm rem}=\int_0^{M_{\rm rem}}2/3\,\Omega\,r^2\,dm$
in units of $\rm 10^{50}\,[ g\,cm^2\,s^{-1}]$,
\item column 4 is the moment of inertia of the remnant, 
${\cal I}_{\rm rem}=\int_0^{M_{\rm rem}}2/3\,r^2\,dm$
in units of $10^{55}\,[{\rm g\,cm}^2]$,
\item column 5 is the specific angular momentum at the remnant edge, $j_{\rm rem}$,
in units of $10^{16}\,[{\rm cm}^2\,{\rm s}^{-1}]$,
\item column 6 is the average specific angular momentum in the remnant, 
$\overline{j}_{\rm rem}={\cal L}_{\rm rem}/M_{\rm rem}$, 
in units of $10^{16}\,[{\rm cm}^2\,{\rm s}^{-1}]$,
\item column 7 is the specific angular momentum at the mass coordinate 1.56 $M_\odot$, 
$j_{1.56}$, in units of $10^{16}\,[{\rm cm}^2\,{\rm s}^{-1}]$, 
\item column 8 is the angular momentum contained in the inner 1.56 $M_\odot$,
${\cal L}_{1.56}=\int_0^{1.56\,M_\odot}2/3\,\Omega\,r^2\,dm$
in units of $10^{50}\,[{\rm g\,cm}^2\,{\rm s}^{-1}]$,
\item column 9 is the specific angular momentum at the mass coordinate 2.5 $M_\odot$, 
$j_{2.5}$, in units of $10^{16}\,[{\rm cm}^2\,{\rm s}^{-1}]$, 
\item column 10 is the angular momentum contained in the inner 2.5 $M_\odot$,
${\cal L}_{2.5}=\int_0^{2.5\,M_\odot}2/3\,\Omega\,r^2\,dm$
in units of $\rm 10^{50}\,[ g\,cm^2\,s^{-1}]$.
\end{itemize}
\begin{figure}[!tbp]
\centering
\includegraphics[width=8.5cm]{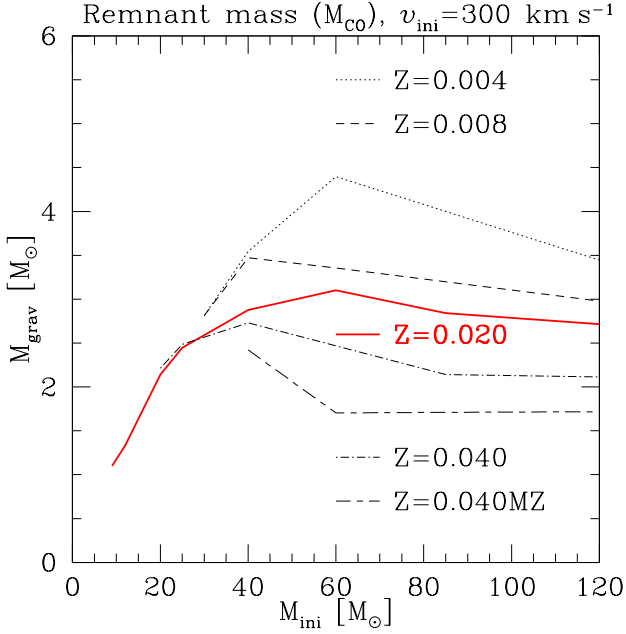}
\caption{
Remnant gravitational masses (estimated from the CO core masses) as a
function of the initial mass of the models. 
}
\label{mimg}
\end{figure}
\begin{table}
\caption{Baryonic masses, $M_b$, gravitational masses, $M_g$ and the BE (all in
$M_\odot$ units, $M_\odot \, c^2=1.79\,10^{54}$erg) of the 
different models calculated using Eq. (\ref{bemg})
and $R=15.12\,M_b^{-1/3}$ \citep[Newtonian polytropes][ p. 245]{ST83}.} \label{tabe}
\scriptsize
\begin{center}
\begin{tabular}{c r r r}
\hline
 \hline \\
 $M_{\rm ini}$   & $M_b$ & $M_g$ & BE   
  \\
 \hline \\
\multispan{4}{\hfill Z=0.004 \hfill } \\ \\
  30     &    3.7284  &  2.8140  &  0.9144 \\
  40     &    5.3954  &  3.5430  &  1.8524 \\
  60     &    8.5150  &  4.3964  &  4.1186 \\
  60$^a$ &    3.7916  &  2.8464  &  0.9452 \\
 120     &    5.1429  &  3.4478  &  1.6951 \\
\\ \multispan{4}{\hfill Z=0.008 \hfill } \\ \\
  30     &    3.7256  &  2.8125  &  0.9130 \\
  40     &    5.2110  &  3.4740  &  1.7370 \\
  60     &    4.9085  &  3.3550  &  1.5535 \\
 120     &    4.0571  &  2.9780  &  1.0792 \\
\\ \multispan{4}{\hfill Z=0.020 \hfill } \\ \\
   9     &    1.1808  &  1.1012  &  0.0796 \\
  12     &    1.4616  &  1.3338  &  0.1278 \\
  15     &    1.8496  &  1.6361  &  0.2135 \\
  20     &    2.5661  &  2.1382  &  0.4280 \\
  25     &    3.0558  &  2.4424  &  0.6135 \\
  40     &    3.8527  &  2.8773  &  0.9754 \\
  60     &    4.3228  &  3.1027  &  1.2201 \\
  85     &    3.7764  &  2.8386  &  0.9377 \\
 120     &    3.5392  &  2.7145  &  0.8247 \\
\\ \multispan{4}{\hfill Z=0.040 \hfill } \\ \\
  20     &    2.6804  &  2.2119  &  0.4685 \\
  25     &    3.1257  &  2.4834  &  0.6423 \\
  40     &    3.5718  &  2.7320  &  0.8399 \\
  85     &    2.5705  &  2.1410  &  0.4295 \\
 120     &    2.5329  &  2.1164  &  0.4165 \\
\\ \multispan{4}{\hfill Z=0.040, $\dot{M}_{\rm WR}(Z)$ \hfill } \\ \\
  40     &    3.0225  &  2.4226  &  0.5999 \\
  60     &    1.9391  &  1.7027  &  0.2364 \\
  85     &    1.9520  &  1.7122  &  0.2398 \\
 120     &    1.9637  &  1.7208  &  0.2429 \\
 \hline
\end{tabular}\\
$^a$ with $\upsilon_{\rm ini}$ = 500 km\,s$^{-1}$                                  
\end{center}
\end{table}
Assuming that a neutron star with a baryonic mass, $M_b= 1.56 M_\odot$, and with a 
 radius, $R=12\,$km, would
form from the models, we calculated the two following quantities:
\begin{itemize}
\item (column 11) the ratio of the NS angular velocity to its 
Keplerian angular velocity, $\Omega/\Omega_K$(NS).
\item (column 12) the initial rotation period of the neutron star,
 ${\cal P}_{\rm rot}$ in units of milli--seconds.
\end{itemize}
\begin{figure*}[!tbp]
\centering
\includegraphics[width=8.5cm]{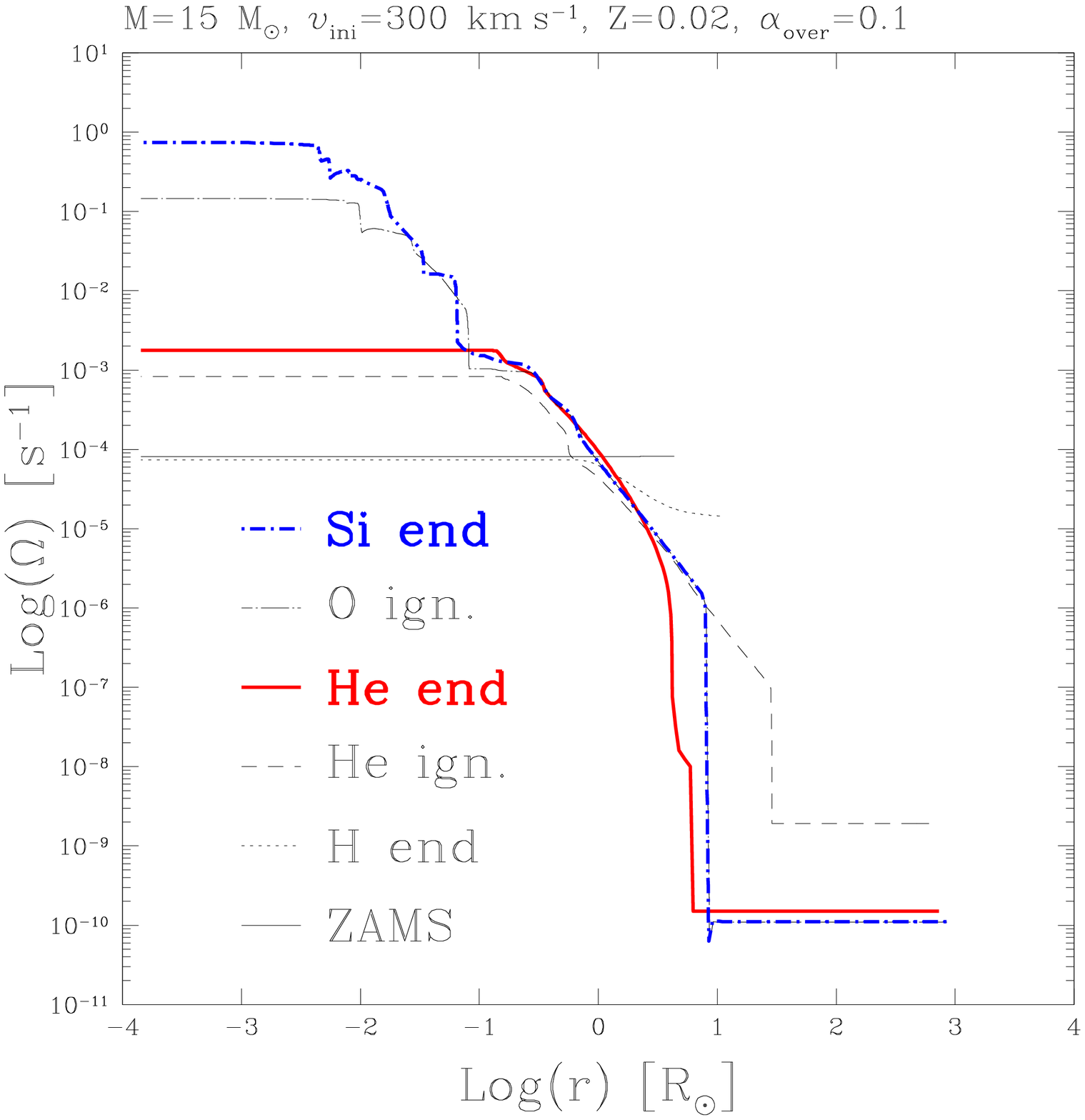}\includegraphics[width=8.5cm]{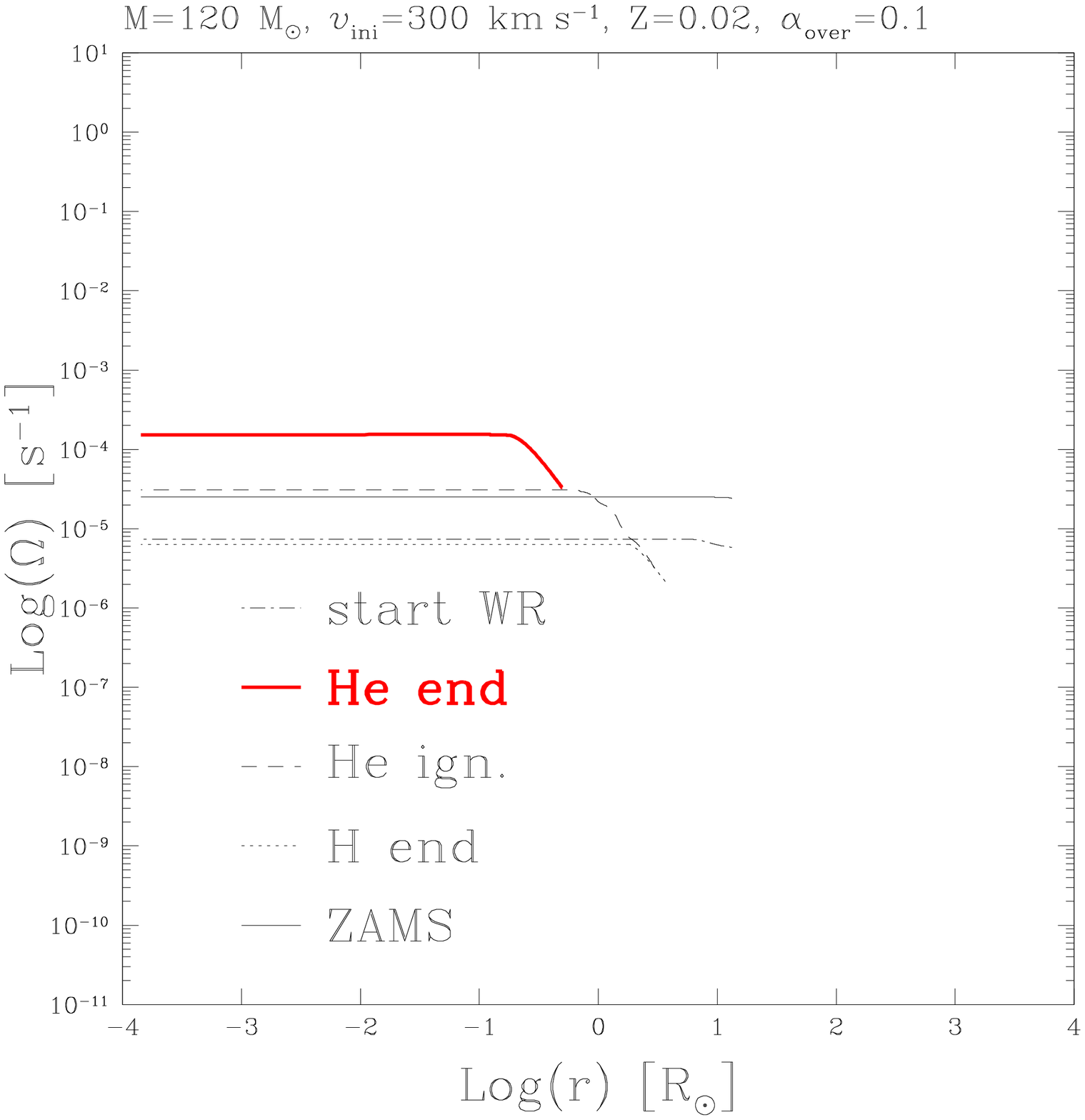}
\caption{Evolution of the angular velocity as a function of the radius at
different stages of the evolution. 
{\it Left:} the 15 $M_\odot$ model becomes and ends up
as a RSG. It therefore has a strong differential rotation between the core and the
envelope. 
{\it Right:} the 120 $M_\odot$ model becomes a WR before the end of the MS and will not go
through the RSG stage. This last model does not have a strong differential rotation
between the core and the envelope and should not be slowed down significantly by magnetic
fields.
}
\label{OWR}
\end{figure*}

A proto--neutron star with a baryonic mass, $M_b= 1.56\,M_\odot$, loses a binding
energy (BE) of 0.159 $M_\odot$ due to neutrinos emission. This value is 
calculated using Eq. (36) from 
\citet{LP01}: BE/$M=0.6\,\beta/(1-0.5\beta)$, 
in which $M=M_g$ corresponds to the gravitational mass 
and where $\beta=G\,M_g/R\,c^2$,
and using the fact that: 
$M_b={\rm BE}+ M_g$. 
These equations give one second degree equation:
\begin{equation}\label{bemg}  
0.1\,G\,M_g^2 - (R\,c^2+0.5\,G\,M_b)M_g - R\,c^2\,M_b=0 
\end{equation}from which we can find the gravitational mass, 
$M_g= 1.401 M_\odot$, which is very close to the average observational 
value of neutron star (gravitational) masses \citep{KM05}.
This is why we chose the value of 1.56 for the baryonic mass. The radius of 12 km
was estimated using Fig. 2 from \citet{LP01}.
The NS momentum of inertia is calculated using Eq. (29) from
\citet{LP01}, which for the mass and radius chosen here corresponds to 
$I_{\rm NS}({\rm T VII})/M\,R^2 \simeq 0.36$. 
Since the core loses mass during the collapse, it also loses angular momentum. 
We assumed here that the angular momentum loss due to neutrino losses
is proportional to the quantity of mass lost. Therefore the angular
momentum of the NS,  
${\cal L}_{\rm NS}={\cal L}_{1.56}\,(M_g/M_b)$. 
This assumes that neutrinos are
emitted uniformly from the entire core. This is probably correct but additional
angular momentum may be lost due to the interaction of these neutrinos with the
outer layers of the core \citep{HWS05}. 

Assuming uniform rotation in the neutron
star and using the relation ${\cal L}_{\rm NS}=I_{\rm NS}\,\Omega_{\rm NS}$, we
have $\Omega_{\rm NS}={\cal L}_{\rm NS}/I_{\rm NS}$.
Taking into account no angular momentum losses other than those due to
neutrinos described above, \\ \\
$\Omega_{\rm NS}$$=$${\cal L}_{1.56}\,(M_g/M_b)/(0.36\,R^2\,M_g)$$=$$
{\cal L}_{1.56}/(0.36\,R^2\,M_b)$.  \\ \\

The Keplerian angular velocity is $\Omega_{\rm K}=\sqrt{G\,M_g/R^3}$. Finally, the
initial period of rotation of the neutron star is given by 
${\cal P}_{rot}=2\,\pi/\Omega_{\rm NS}$.

We also give values of the angular momentum at and 
in the inner 2.5 $M_\odot$ (near maximum mass for neutron stars).
These results allow to
compare different models at a given mass coordinate rather than at the 
remnant edges which span a wide range of masses (see Tables \ref{moma04}--\ref{moma40}).
Finally, we used Eq. (\ref{bemg}) to obtain
the gravitational masses, $M_g$ and the BE 
(all in $M_\odot$ units, $M_\odot \, c^2=1.79\,10^{54}$erg) 
for our models
using $R=15.12\,M_b^{-1/3}$ \citep[Newtonian polytropes][ p. 245]{ST83}
These values are given in Table \ref{tabe} and the gravitational masses are shown in
Fig. \ref{mimg}. We see that, using our estimates, black holes are expected to form from
stars with masses above about 20 $M_\odot$ for solar or lower metallicities. At higher
metallicities, the formation of a black hole or neutron star depends on the mass loss
prescription. At Z=0.040 and using a mass loss dependent on metallicity (models MZ), 
we expect black hole to form from stars with masses between about 20 and 55 $M_\odot$.

\section{Evolution of rotation}

\begin{figure*}[!tbp]
\centering
\includegraphics[width=8.5cm]{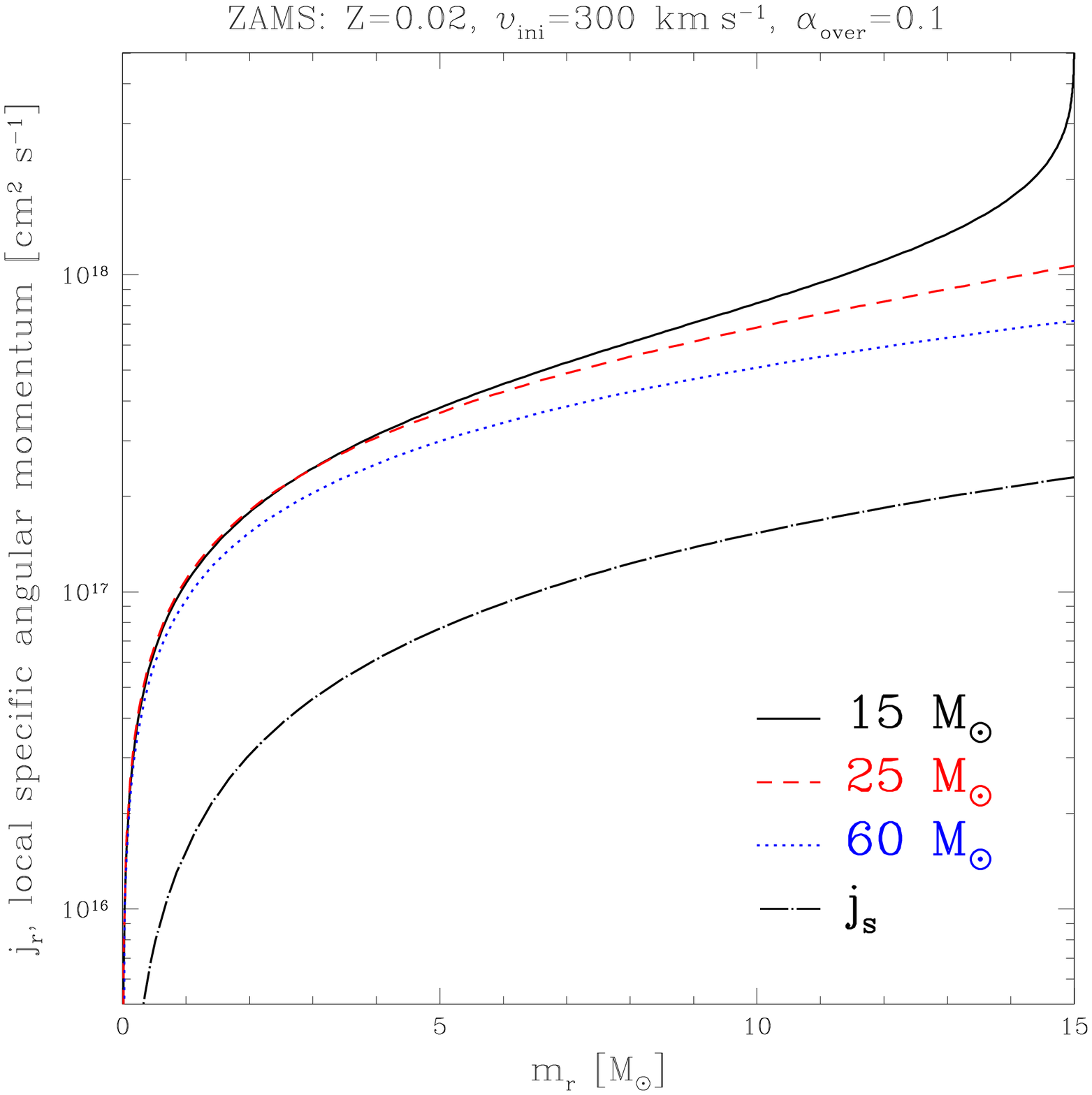}\includegraphics[width=8.5cm]{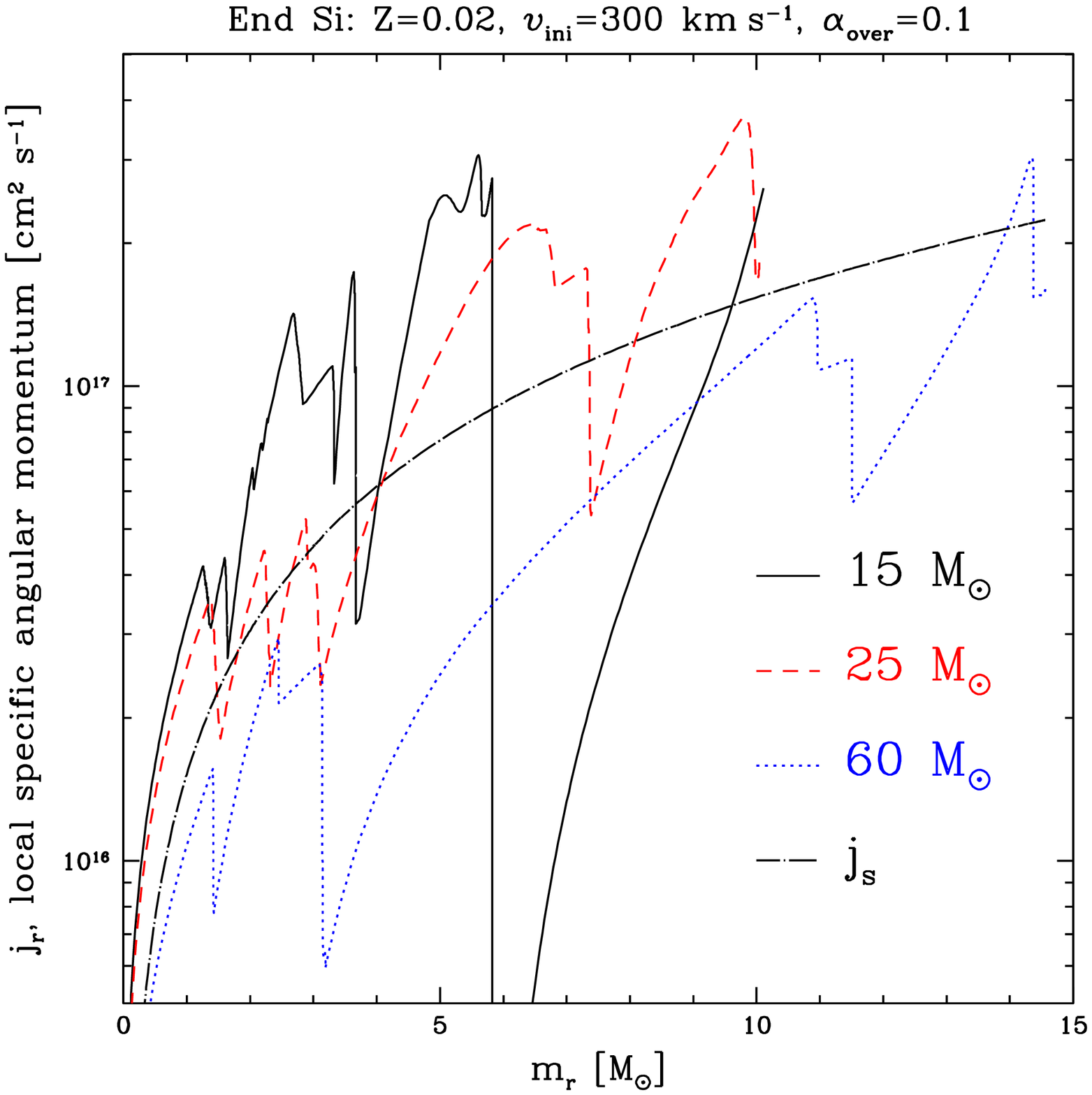}
\caption{Profiles of the specific angular momentum, $j_r$, as a function
of the Lagrangian mass coordinate, $m$, for different models:
15 $M_\odot$ (solid lines), 25 $M_\odot$ (dashed lines) and 
60 $M_\odot$ (dotted lines). 
{\it Left:}  On the ZAMS.
{\it Right:} At the end of Si--burning.
$j_S=\sqrt{12}G\,m/c$ (dashed--dotted line) is the specific angular momentum 
necessary for matter to form an accretion
disk around a non--rotating black hole (Schwarzschild metric) and is displayed
here as a reference profile.
The specific angular momentum varies significantly with the mass coordinate at
the end of Si--burning. For the average profile, the heavier the initial mass of 
the star, the
stronger the mass loss and angular momentum loss. Therefore the core of the 60
$M_\odot$ model ends with much less angular momentum than on the ZAMS and
loses more angular momentum in comparison with lighter models (15, 25 $M_\odot$).
}
\label{jcsol}
\end{figure*}

In paper XII, we presented the evolution of the angular velocity profiles
through the pre-supernova stages for the solar metallicity 25 $M_\odot$ model.
We showed that the angular velocity increases in the core via successive 
contractions. In the outer part of the star, after the MS, $\Omega$ either has very
small values,
when the star is a red supergiant (RSG), or has values similar to the initial
values when the H--rich envelope of the star is lost and the star becomes a WR star. 
If a star is
very massive, the star may become a WR before the end of the main sequence (MS) and skip
the RSG phase (see Fig. \ref{OWR}). 
This point has important consequences for the braking of
stars by magnetic fields, as mentioned in \citet{HWS05}. Indeed in the theory
elaborated by \citet{Sp02}, differential
rotation produces strong magnetic fields, which in turn slow down the core of the
star. This means that stars that go through the RSG phase will experience a 
strong braking due to the large differential rotation between the core and the envelope
while stars that skip the RSG stage might be slowed down much less.

The evolution of the angular momentum profiles has also been discussed in 
paper XII. In our models, we consider three transport processes: convection,
shear diffusion and meridional circulation. Convection makes $\Omega$ constant
and therefore transports angular momentum from the inner part of a convective
zone to the outer part of the same convective zone. Shear diffusion reduces
differential rotation and also transports angular momentum outwards. 
Meridional circulation can transport angular
momentum inwards or outwards. Mass loss acts somewhat 
indirectly on the angular momentum of the core 
by affecting $\Omega$ and the gradients of $\Omega$ inside the 
star and thus the efficiency of the transport processes. 
The main conclusions about the evolution of the angular momentum are the 
following:

- The angular momentum generally decreases throughout the pre--supernova
evolution. 

- The largest decrease occurs during the MS.

-  After He--burning, only convective
zones reshape the angular momentum profile and produce teeth along its 
profile 
(redistributing angular momentum from the inner boundary of a convective 
zone to 
the outer boundary of the same zone) without
removing significant amount of angular momentum from the core.
This shows that the angular momentum at the end of He-burning is a good 
approximation of
the angular momentum contained in the core at the collapse.
\begin{figure}[!htbp]
\centering
\includegraphics[width=8.5cm]{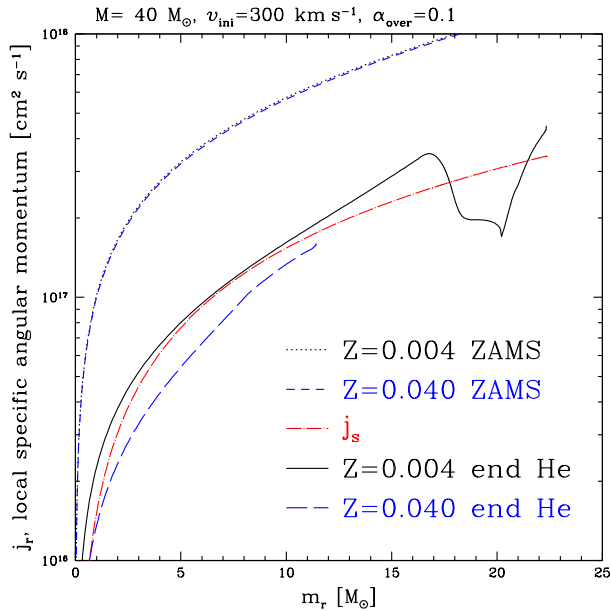}
\caption{Profiles of the specific angular momentum, $j_r$, as a function
of the Lagrangian mass coordinate, $m$, at
the ZAMS and at the end of He--burning  
for 40 $M_\odot$ models at Z=0.004 and Z=0.040.
The metallicity dependence of the mass loss affects the evolution of the profile,
especially at the surface. The decrease of angular momentum is larger at
higher metallicity due to a larger mass loss and a more efficient transport.
The specific angular momentum necessary to form a accretion disk around
a non--rotating black hole (Schwarzschild metric), 
$j_{\rm S}=\sqrt{12}\,G\,M/c$ is given for reference. We see that a low
metallicity star is more likely to form an accretion disk than a high
metallicity star.
}
\label{jevol40}
\end{figure}

\subsection{Dependence on the initial mass}
The variation of both the remnant mass and the specific angular momentum as
a function of the mass coordinate makes the comparison between different models quite
difficult at the pre--supernova stage (see Fig. \ref{jcsol}).
Indeed, the specific angular 
momentum, $j$, varies significantly in the core (see Fig. 7 in paper XII). 
Until the end of
He--burning, $j$ increases monotonically with the mass coordinate. 
At the end of
Si--burning, the teeth produced by the convective zones make the final $j$ at a
given mass coordinate oscillate between values slightly above or a few times
under the values at the end of He--burning. 
The remnant mass also obviously varies with the initial mass (see values in Tables 
\ref{moma04}--\ref{moma40}).

This is why we also give the values of the specific angular momentum at 
fixed Lagrangian mass coordinates (1.56 and 2.5) 
and give the average value inside the remnant (${\overline j}_{\rm rem}$).
It is easier to compare between models at the end of He--burning and this is what
we do in the next section. Nevertheless, we present the values at the end of
Si--burning for solar metallicity models to show how the specific angular 
momentum at a given mass coordinate can vary during the advanced stages.

The general dependence on the initial mass of the final specific
angular momentum in the core can be seen in Fig. \ref{jcsol}.
The larger the initial mass, the smaller the final specific angular momentum in the
core. This is due to a stronger mass loss and a more efficient transport of
angular momentum out of the core. 
The more efficient transport in more massive stars can be explained by the
smaller compactness of these stars. Indeed, higher compactness reduces the 
outward transport by
meridional circulation (Gratton--\"Opik term proportional to $\rho^{-1}$, 
see paper VII).

\subsection{Dependence on the metallicity}
Figure \ref{jevol40} shows the initial and final angular momentum profiles (as a
function of the Lagrangian mass coordinate) for 40 $M_\odot$ models at
metallicities equal to 0.004 (SMC) and 0.040 (Galactic
center). 
Since
higher metallicity implies stronger mass loss and less compact stars, the
dependence of the final specific angular momentum in the core on the metallicity
is similar to the dependence on the initial mass of the star. Therefore, the
higher the metallicity, the smaller the final angular momentum in the core.

\section{Models at the final stages and predictions for 
long soft gamma ray bursts (GRBs)}
Stellar mass black hole candidates (Cyg X--3, LMC X--1, ...) 
are mostly observed in X--ray binaries \citep[see][ and references
therein]{KM05}. 
Usually, a lower limit for the
mass of the accreting body which is larger than about 2 to 3 solar masses makes
it a black hole candidate. 
Observationally derived masses range from about 3 to 15 
(for GRS1915+105).
The gravitational masses in Fig. \ref{mimg} (and Table \ref{tabe}) can give us a 
very rough idea of which models may form BHs.
Black holes are expected to form from
stars with masses above about 20 $M_\odot$.
The upper mass limit for BH formation depends on the mass loss
prescription. At Z=0.040 and using a mass loss dependent on metallicity (models MZ), 
we expect black hole to form from stars with masses between about 
20 and 55 $M_\odot$ (consdering models with remnant masses larger than 
2 M$_\odot$ in Fig. \ref{mimg}).
 
\begin{figure*}[!tbp]
\centering
\includegraphics[width=8.5cm]{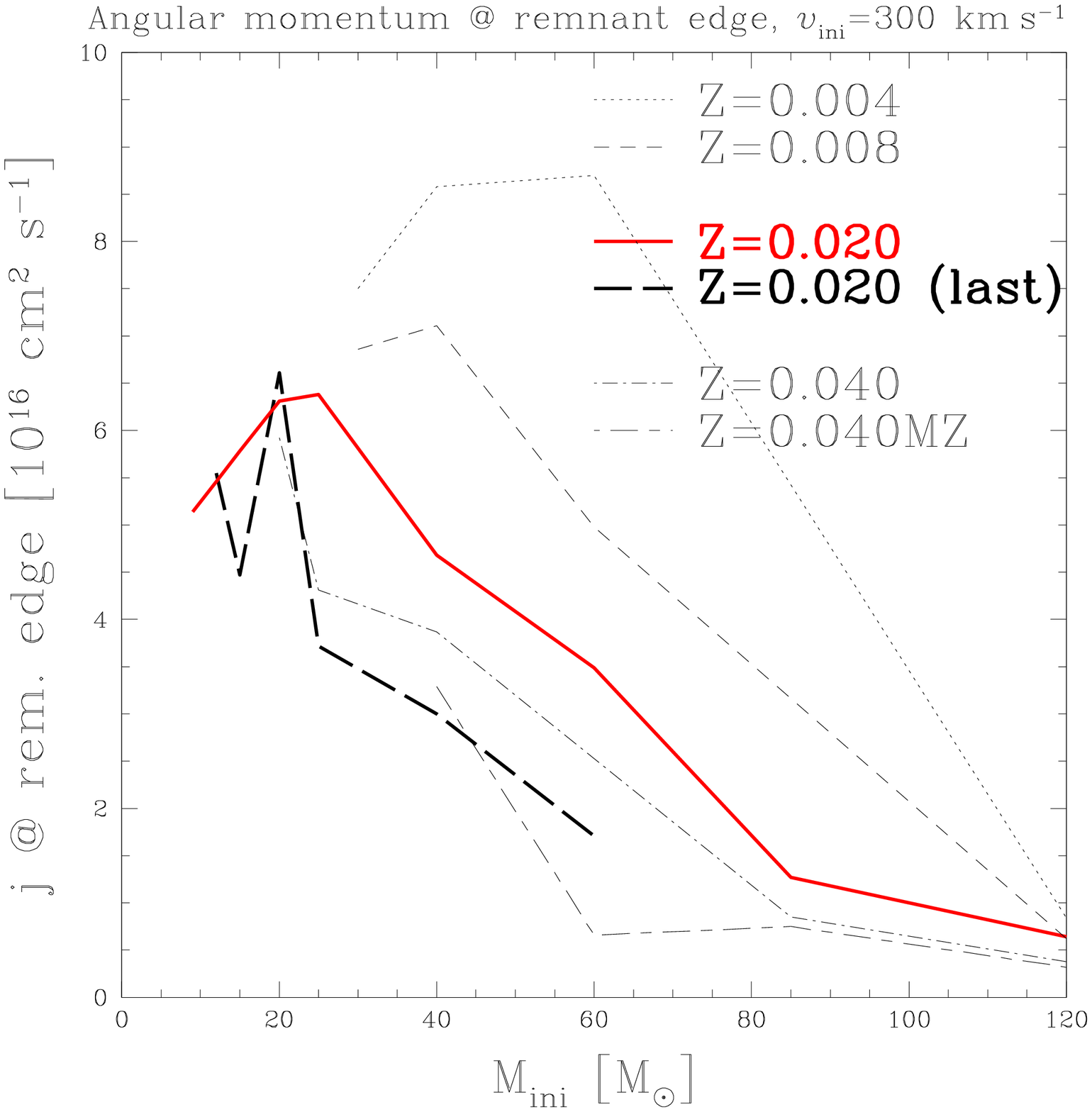}\includegraphics[width=8.5cm]{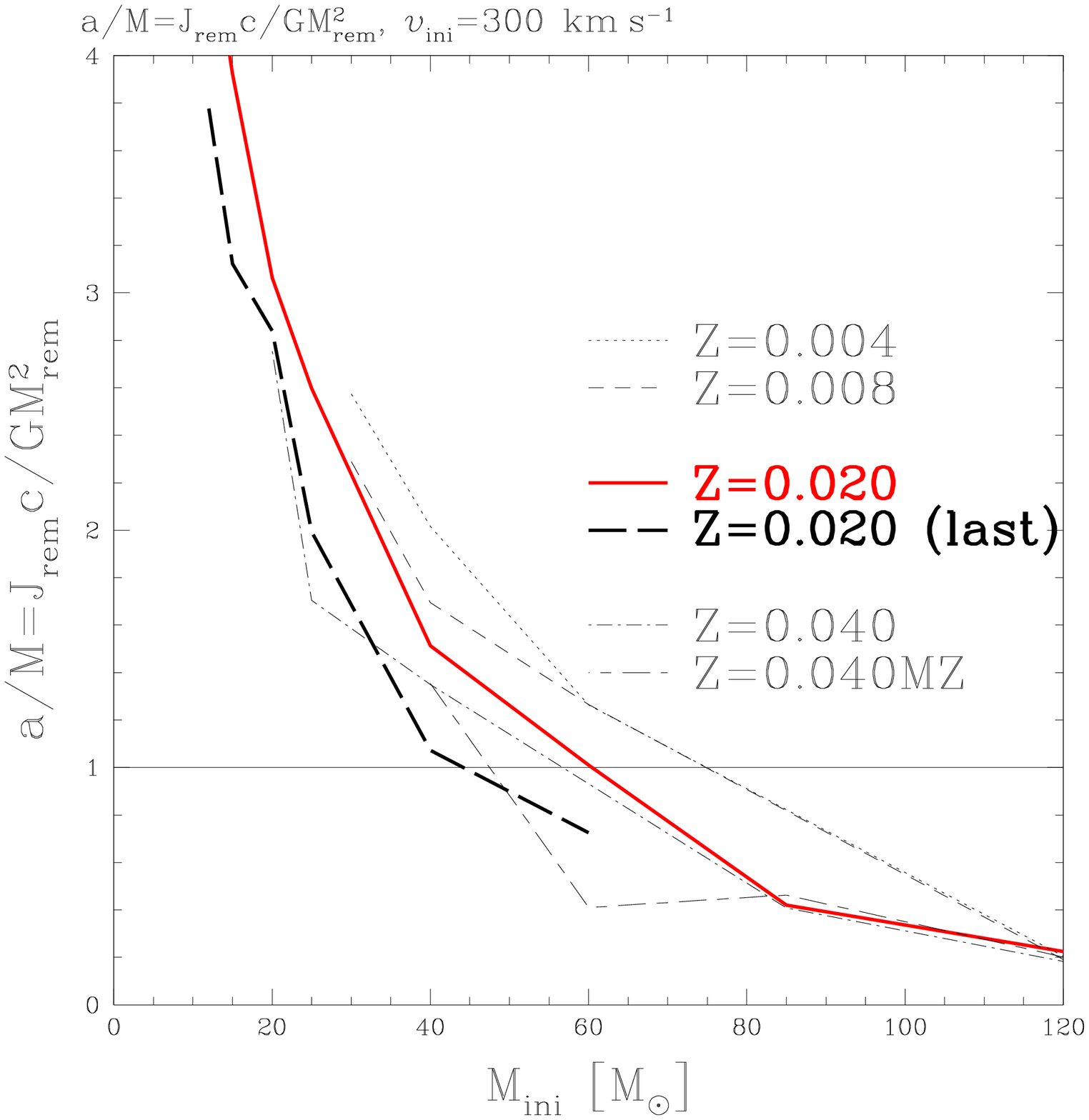}
\caption{
{\it Left:} Specific angular momentum at the edge of the remnant as a
function of the initial mass of the models. The values are determined
from the models at the end of He--burning. For solar metallicity models,
values determined at the end of Si--burning (last) are also given.
Short--long dashed lines represent models with Z=0.040 with a mass loss
during the WR phase which depends on the metallicity 
\citep[MZ, see][]{ROTXI}.
At the edge of the remnant, values at the end of Si--burning are usually 
lower than at the end of
He--burning but they can also be slightly larger as is the case for the
20 $M_\odot$ model here.
{\it Right:} Angular momentum per unit mass, $a=J\,c/G\,M$, divided by
the mass of the remnant, $M$, as a function of the initial mass of the
models. The horizontal line with $a/M=1$ corresponds to the maximally 
rotating BHs.
}
\label{grb1}
\end{figure*}

No observation of BH rotation is
available. Theoretically, one can use the Kerr metric to describe 
rotating black
holes. In this metric, the parameter $a=J\,c/G\,M$ denotes the angular
momentum per unit mass (in units of gram if the other quantities are in
CGS units: $G$, the gravitational constant, equals 
$\rm 6.67\,10^{-8}\,cm^3\,s^{-2}\,g^{-1}$,
$c$, the speed of light, equals $\rm 2.9979\,10^{10}\,cm\,s^{-1}$,
$M$, the remnant mass, is in grams
and $J$, the angular momentum of the remnant is in $\rm g\,cm^2\,s^{-1}$).
In the Kerr metric,
$a/M$ can take values between 0 (non--rotating BH:
the Kerr metric becomes equivalent to the Schwarzschild metric)  
and 1 (1 for maximally rotating BHs).
Figure \ref{grb1} (right) shows the predicted values for $a/M$, assuming no loss
of angular momentum during collapse (apart from a uniform loss due to
neutrino emission). Stars lighter than about 20 $M_\odot$ probably form
neutron stars. We see that models between 20 and 60 $M_\odot$ have 
values exceeding the maximal value. These models probably form an accretion
disk before the black hole is fully formed.

Accretion disks around BHs are very interesting in the context of long
soft GRBs. These bursts were recently connected with SNe of type Ib,c
\citep[see][ for example]{Ma03}.
Three current scenarios \citep{WH03} for GRB production are
the supranovae model \citep{VS99}, 
the magnetar model \citep{WMW02} and 
the collapsar mechanism \citep{W93}. In this last mechanism, a star
collapses into a black hole and an accretion disk due to the high
angular momentum of the core. Accretion from the disk onto the central
black hole produces bi--polar jets. These jets can only reach the
surface of the star (and be detected) if the star loses 
its hydrogen rich envelope before the collapse. 
WR stars are therefore good candidates for collapsar
progenitors since they lose their hydrogen rich envelope during the
pre--SN evolution. Which WR stars form BHs and which have enough angular
momentum to form an accretion disk around them? The first part of the
question can be answered by looking at the estimated 
gravitational mass of the remnants presented in Fig. \ref{mimg} and 
given in Table \ref{tabe}.
In this study, using our rough estimate for the remnant mass, 
the minimum initial mass for black hole
formation is between 20 $M_\odot$  
(2 $M_\odot$ as the maximum mass for neutron stars) and
around 30 $M_\odot$ (2.5 $M_\odot$).
We use for GRB rate predictions the lower limit of 20 $M_\odot$. 
Note for the following discussion that the present WO models definitely 
form BHs.
In order to form an accretion disk around a BH, the matter on the last
stable orbit (LSO) must have an angular momentum larger than 
$j_K=r_{\rm LSO}\,c$ \citep[ p. 428]{ST83}.
The radius of the last stable orbit, $r_{\rm LSO}$, is given by 
$r_{\rm ms}$ in formula
(12.7.24) from \citet[p. 362]{ST83} for circular orbit in the Kerr
metric. 
$r_{\rm LSO}$ depends on $a$ and the direction of the orbit. We 
consider here 
direct (co--rotating) orbits. In this case $r_{\rm LSO}$ varies from 
$G\,M/c^2$ for $a=1$ to $5\,G\,M/c^2$ for $a=0$.
Note that, for the Schwarzschild metric, $r_{\rm LSO}=\sqrt{12}\,G\,M/c^2$
and $j_S=\sqrt{12}\,G\,M/c$.
For models with $a>M$, we used $r_{\rm LSO}=G\,M/c^2$.
\begin{figure*}[!tbp]
\centering
\includegraphics[width=8.5cm]{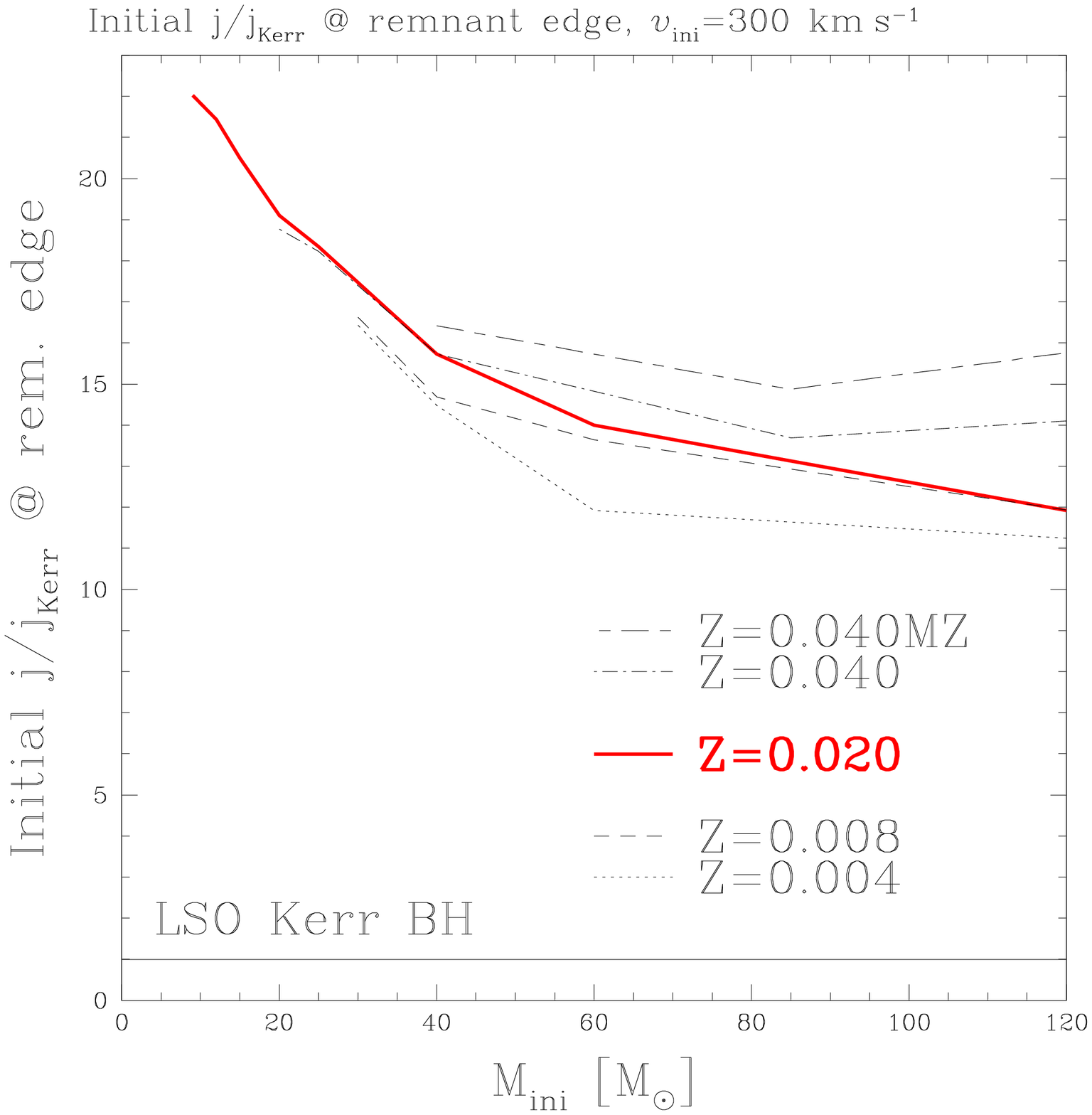}\includegraphics[width=8.5cm]{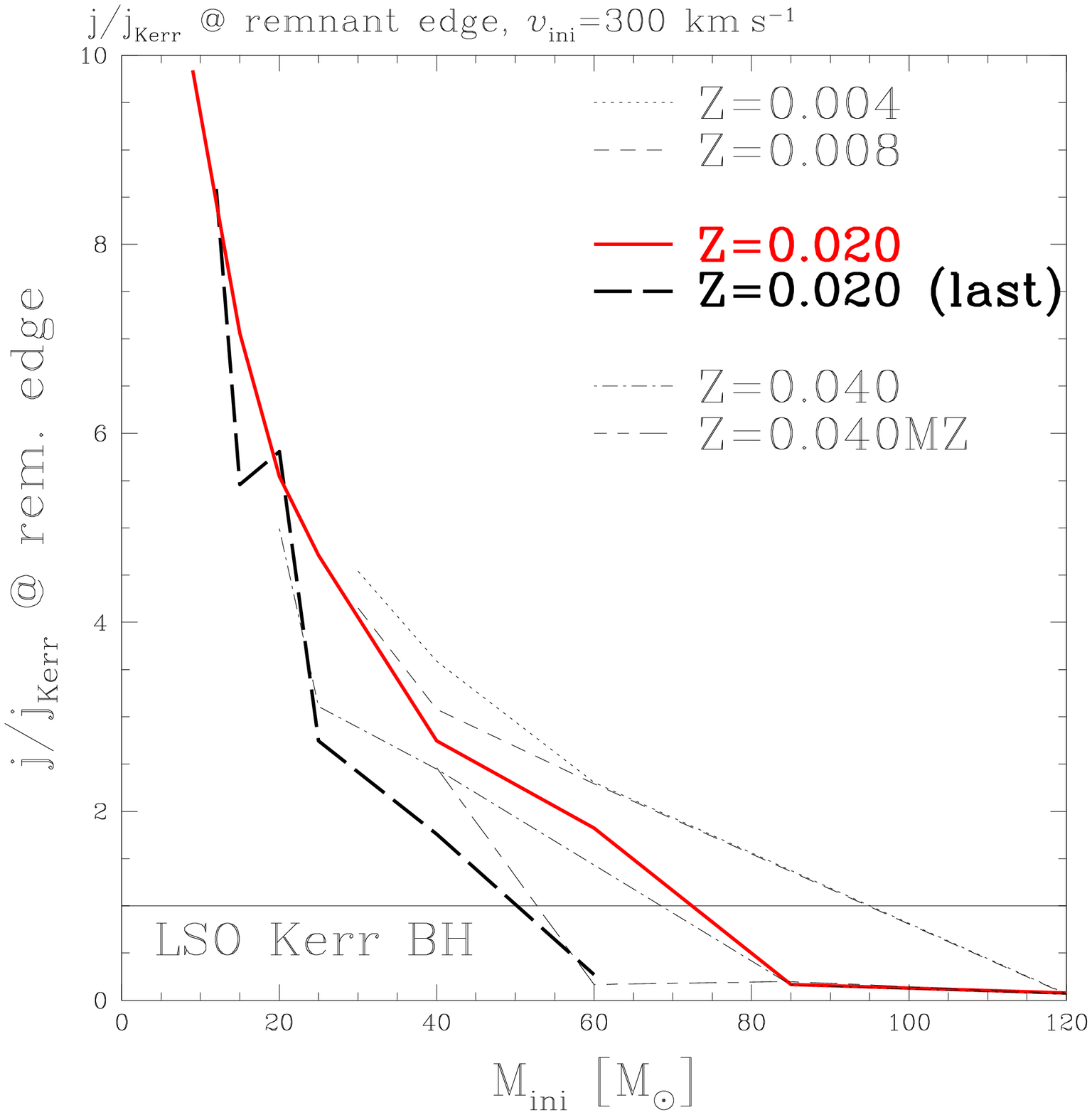}
\includegraphics[width=8.5cm]{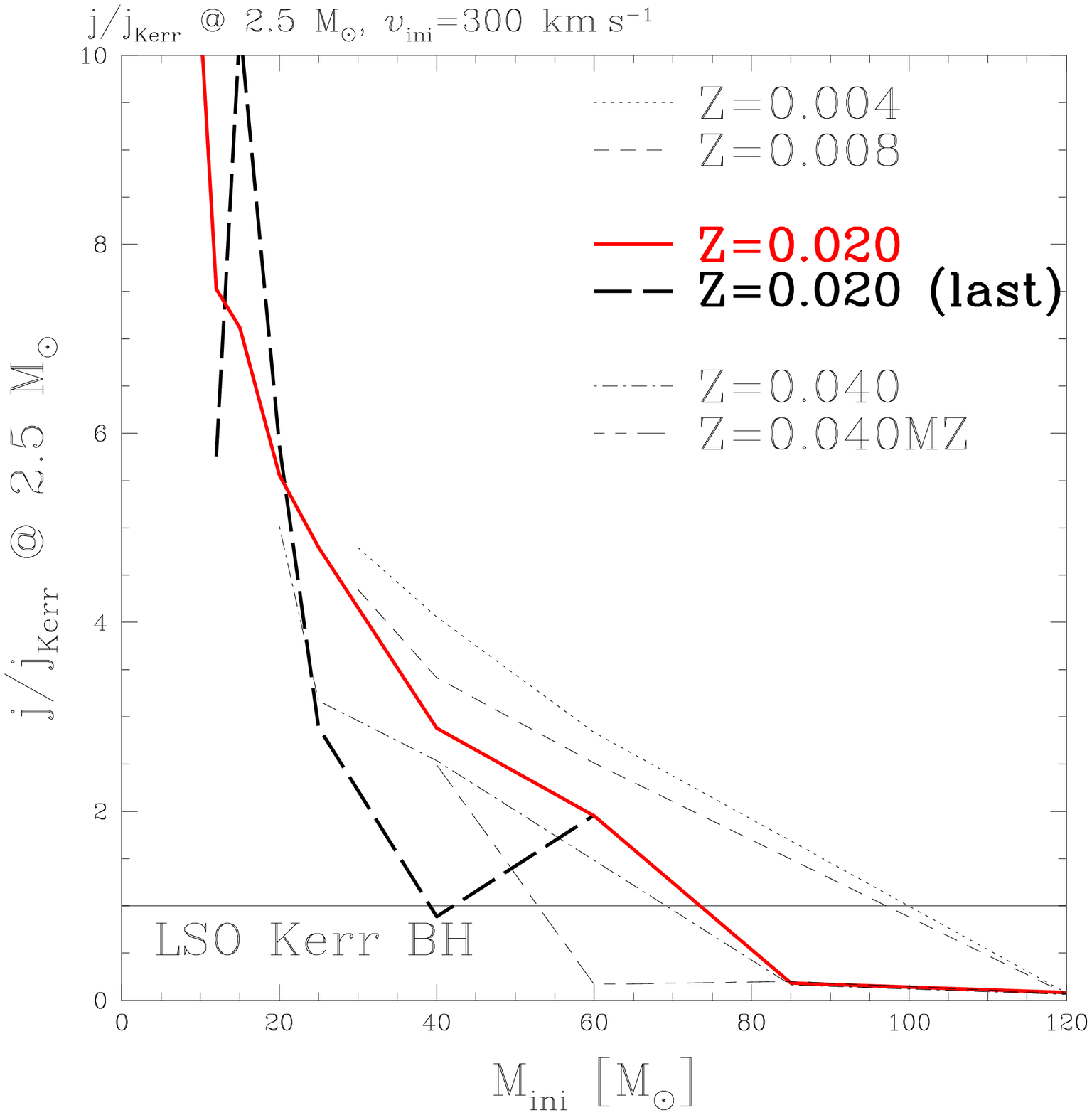}\includegraphics[width=8.5cm]{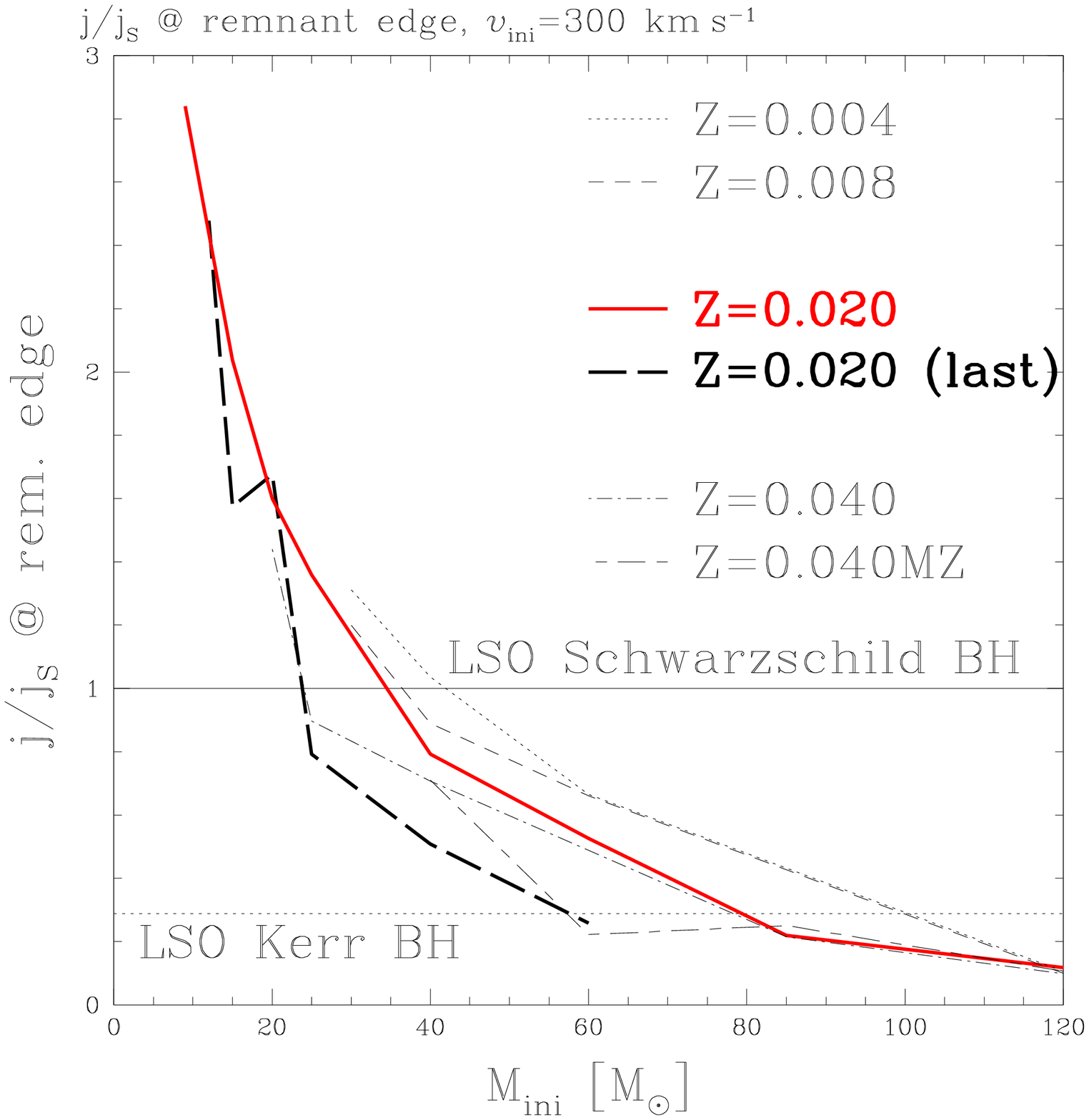}
\caption{
{\it Top left:} Initial ratio of the specific angular momentum at the edge of 
the remnant to $j_K=r_{\rm LSO}\,c$ (the value of angular momentum
necessary to form an accretion disk at the edge of the remnant) on the ZAMS as a
function of the initial mass of the models. The initial ratio is shown
to see whether trends observed at the end are due to initial conditions
or differences in the evolution.
{\it Top right:} Final ratio of the specific angular momentum at the edge of 
the remnant to $j_K=r_{\rm LSO}\,c$ as a
function of the initial mass of the models. The values are determined 
from models
at the end of He--burning. At solar metallicity, values at the end of
Si--burning (last) are also given.
{\it Bottom left:} Ratio of the specific angular momentum at the
Lagrangian mass coordinate of 2.5 $M_\odot$ to $j_K=r_{\rm LSO}\,c$ as a
function of the initial mass of the models. This graph is used to compare
models at a fixed Lagrangian mass coordinate.
{\it Bottom right:} Ratio of the specific angular momentum at the edge of 
the remnant to $j_S=\sqrt{12}\,G\,M/c$ as a
function of the initial mass of the models. In this
 $j/j_S$ plot, the maximally rotating BH value of $j_K/j_S=1/\sqrt{12}$ is
 also given for reference. 
Short--long dashed lines represent models with Z=0.040 with a mass loss
during the WR phase which depends on the metallicity 
\citep[MZ, see][]{ROTXI}.
At the edge of the remnant, values at the end of Si--burning are usually 
lower than at the end of
He--burning but they can also be slightly larger as is the case for the
20 $M_\odot$ model here.
}
\label{grb_jk}
\end{figure*}
Figure \ref{grb1} (left) shows the specific angular momentum at the edge
of the remnant, $j_{\rm rem}$, for the different models. The values are derived from
models at the end of He-burning. For solar metallicity models,
values determined at the end of Si--burning are also given.
At the edge of the remnant, the specific angular momentum 
at the end of Si--burning is usually smaller than at the end of
He--burning but it can also be slightly larger as is the case for the
20 $M_\odot$ model here. We therefore use values at the end of
He--burning for our discussion here.

Between 15 and about 40 $M_\odot$, $j_{\rm rem}$ increases 
(see Fig. \ref{grb1}). This is due
to the increase in the remnant size and to the fact that the specific
angular momentum increases with mass coordinate (see Fig.
\ref{jevol40}). If we look at the same mass range in Fig. \ref{grb_jk},
we see that the lower mass stars rotate faster with respect 
to break--up rotation than heavier
stars. In fact, the trend is monotonic: the heavier the star, the
slower (compared to the maximally rotating BHs) it rotates.
This is true not only at the remnant edge but also at different mass
coordinates: 1.56 $M_\odot$ (see Tables \ref{moma04}--\ref{moma40}) and 2.5
$M_\odot$ (see Fig. \ref{grb_jk} bottom--left).
Even though this trend is already present on the ZAMS 
(see Fig. \ref{grb_jk} top--left), it is much weaker and the
evolution is responsible for the final results. As said earlier, the
stronger mass loss and more efficient transport of angular momentum are
responsible for the lower final angular momentum in the core at high metallicities.
For the metallicity dependence, Fig. \ref{grb_jk} (top--left) 
shows that lower metallicity models with a same initial mass 
start with a smaller ratio
$j/j_{\rm K}$ (and end with a higher one). 
The lower initial ratio is mainly due to the different (higher) 
remnant mass. 
Indeed, $j_{\rm K}$ is proportional to the remnant mass and
$j=\Omega\,r^2$ is roughly proportional to $M_{\rm rem}^{2/3}$ 
(if $\Omega$ and $\rho$ are constant), which means that
the ratio $j/j_{\rm K}$ is roughly proportional to $M_{\rm rem}^{-1/3}$.
Lower metallicity models having larger remnant masses start with a
smaller ratio.
In this study we choose the same initial
surface velocity for models of different compactness 
(lower metallicity = higher
compactness) and not the same angular momentum or same ratios 
$j/j_{\rm K}$ or $\Omega/\Omega_{\rm c}$. Future studies will consider 
models with similar angular momentum or ratios as well 
as identical velocity, which is very important for
extremely low metallicity stars. 
Even if high metallicity models begin with a higher ratio, 
they end with a smaller one. 
Therefore, the trend of higher metallicity leading 
to smaller final angular momentum (compared to the break--up value) 
is confirmed.

Concerning the criterion of angular momentum for GRB production, our
models show that, the lower the initial mass, the more probable the GRB
production. The specific angular momentum necessary to form an accretion disk
around a BH depends on the rotation of the BH itself.
If we want to remove this dependence, we can compare
$j_{\rm rem}$ with the Schwarzschild metric $j_S=\sqrt{12}\,G\,M/c$ 
(see Fig. \ref{grb_jk} bottom--right) although 
a rotating collapsing core must form a rotating black hole. In this
 $j/j_S$ plot, the maximally rotating BH value of $j_K/j_S=1/\sqrt{12}$ 
is given for reference.
This plot shows that models below about 40 $M_\odot$ even have $j/j_S$
values larger than one.
\subsection{Predicted rates of long soft gamma ray 
bursts (GRBs) from single star progenitors}

\begin{table}
\caption{Predicted rates [yr$^{-1}$] of WR stars and of 
long soft gamma ray bursts (GRBs) from single star progenitors
as well as limiting masses at various metallicities.} \label{grbrates}
\scriptsize
\begin{tabular}{l r r r r}
\hline
 \hline \\
\multispan{5}{\hfill Reference rate: 
${\cal R}^{\rm OBS}_{\rm SN}\simeq 7\,10^{-3}$ \hfill } \\ \\
 \hline \\
& $Z_{\rm SMC}$   &   
  $Z_{\rm LMC}$   &   
  $Z_{\odot}$     &   
  $Z_{\rm GC}$       
  \\
 \hline \\
%
${\cal R}_{\rm WR}$  & 1.31E-03 & 1.90E-03 & 2.29E-03 & 2.45E-03 \\
${\cal R}_{\rm WO}$  & 6.33E-04 & 7.58E-04 &    0     &    0     \\
 \hline \\
$M^{\rm min}_{\rm GRB}$(WR) & 32 & 25 & 22 & 21 \\
$M^{\rm max}_{\rm GRB}$(WR) & 95 & 95 & 75 & 55 \\
${\cal R}_{\rm GRB}$(WR) & 1.15E-03 & 1.74E-03 & 2.01E-03 & 1.92E-03 \\
 \hline \\
$M^{\rm min}$(WO) & 50 & 45 & -  & - \\
$M^{\rm max}$(WO) & 110 & 100 & -  & - \\
${\cal R}_{\rm WO}$  & 6.33E-04 & 7.58E-04 &    0     &    0     \\
 \hline \\
$M^{\rm min}_{\rm GRB}$(WO) & 50 & 45 & -  & - \\
$M^{\rm max}_{\rm GRB}$(WO) & 95 & 95 & -  & - \\
${\cal R}_{\rm GRB}$(WO) & 4.74E-04 & 5.99E-04 & -  & -  \\
 \hline \\
\multispan{5}{\hfill   
 ${\cal R}^{\rm OBS}_{\rm GRB} = 3\,10^{-6}-6\,10^{-4} \,^a$ \hfill } \\ \\
 \hline \\
\end{tabular} \\
$^a$ Depending on the beaming angle of GRBs.                        
\end{table}
\begin{table}
\caption{Final WR type for WC/WO stars at various metallicities. 
At $Z_{\rm GC}$, models with metallicity dependent mass loss rates were
used.}
\label{WCWO}
\scriptsize
\begin{tabular}{l r r r r}
\hline
 \hline \\
& $Z_{\rm SMC}$   &   
  $Z_{\rm LMC}$   &   
  $Z_{\odot}$     &   
  $Z_{\rm GC}$       
  \\
 \hline \\
%
25/30 $M_\odot$& -$^a$ & WC6 &-$^a$& WC5 \\
 40 $M_\odot$  & -$^a$ & WC4 & WC4 & WC4 \\
 60 $M_\odot$  & WO    & WO  & WC4 & WC7 \\
 85 $M_\odot$  & ?$^b$ &?$^b$& WC4 & WC6 \\
120 $M_\odot$  & WC4   & WC4 & WC4 & WC7 \\
 \hline \\
 \hline \\
\end{tabular} \\
$^a$ "-" means the model does not become a WC or WO star.\\
$^b$ "?" means the model was not calculated.\\
                        
\end{table}
Following the method used in \citet{PMN04}, 
we present, in Table \ref{grbrates}, predictions for the rate of long 
soft GRBs produced
in the collapse of massive single stars.
For reference, we use the following SN rate: 
${\cal R}^{\rm OBS}_{\rm SN}= c_{\rm SFR}\,
\int_{10 M_\odot}^{150 M_\odot} m^{-2.35}\,dm = 7\,10^{-3}\, {\rm yr}^{-1}$
$(=c_{\rm SFR}\,[150^{-1.35}-10^{-1.35}]/-1.35)$. 
The value of $7\,10^{-3}\, {\rm yr}^{-1}$ is the 
same as in \citep{PMN04}
and represents the total number of core collapse SNe per year for an
average galaxy 
\citep[the Milky Way galaxy has a slightly higher value 
$\sim 1.2\,10^{-2}\, {\rm yr}^{-1}$, see][]{Ca99}. $c_{\rm SFR}$ is a normalisation factor taking
into account the star formation rate (SFR). It is determined by the
equation above. We use Salpeter's IMF \citep{Sa55} and an upper mass limit of
150 $M_\odot$ following the recent results \citep[see][and references
therein]{Kr05}.
The other rates are then calibrated to this reference rate:
for example, 
${\cal R}_{\rm GRB}$(WR)$= (7\,10^{-3}
/ \int_{10 M_\odot}^{150 M_\odot} m^{-2.35}\,dm)
\,^. \int_{M^{\rm min}_{\rm GRB}({\rm WR})}^{M^{\rm max}_{\rm GRB}({\rm WR})} 
m^{-2.35}\,dm 
$.
The rates of WR stars are derived using the minimum mass limits
given in Table 2 from \citet{ROTXI}. 

These minimum masses correspond approximately to the
minimum mass for GRB production since these stars have lost their
hydrogen envelope by definition and probably form BHs 
(have remnant gravitational masses 
$\gtrsim 2\,M_\odot$; see Fig. \ref{mimg} and Table \ref{tabe}).
${\cal R}_{\rm GRB}$(WR) corresponds to the rate of GRBs produced by
massive single stars which form BHs, lose their hydrogen rich envelope
and have enough angular momentum at the edge of the remnant to form an
accretion disk. We can see that the rates are much higher (10--1000 times)
than the observed long soft GRB rate, 
${\cal R}^{\rm OBS}_{\rm GRB}\simeq 3\,10^{-6}-6\,10^{-4}$ \citep{PMN04},
for beaming angles between 15 and 1$^o$.

An additional constraint for GRB production might be that the star must be a
WR of type WO at the time of the explosion. WO stars are WR stars which have
lost not only their hydrogen rich envelope but also the helium rich layers.
They explode as type Ic supernovae.
Although there is no firm
evidence to support this fact, the link between GRBs and
supernovae was firmly established for the type Ic hypernova SN2003dh 
\citep{MMN03}. Furthermore, WO stars have lost both their hydrogen and
helium rich layers. This means that their envelope is more tenuous, 
thus favoring the escape of the
jets. \citet{SM91} studied the metallicity dependence of WR stars of type
WC and WO. They found that the lower the initial metallicity $Z$,
the higher the (C+O)/He number ratio when the star first becomes a WC
star. 
This is due to the fact that at low metallicity, mass loss is
smaller and the newly synthesised C and O are revealed later in the
evolution. It implies that WO stars are mainly produced at low
metallicity. 
\begin{figure}[!tbp]
\centering
\includegraphics[width=8.5cm]{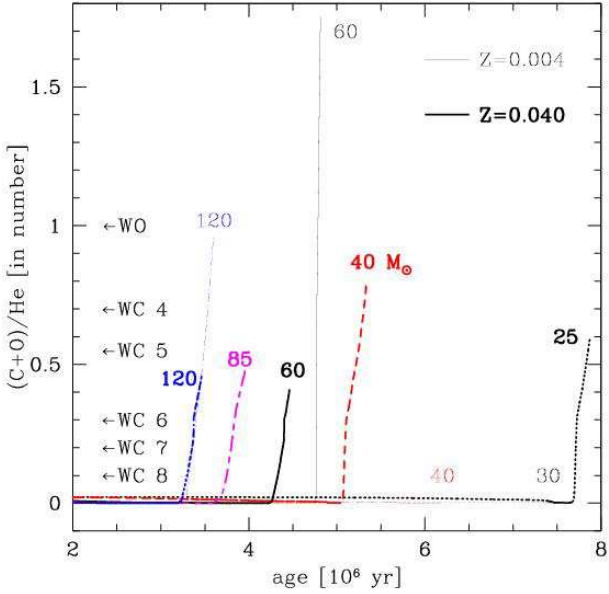}
\caption{Evolutionary tracks in the surface (C+O)/He number ratio 
as a function of age for models with different initial masses
at Z=0.004 (thin lines) and Z=0.040 (thick lines).
The different WC/WO types are indicated. At Z=0.040, most stars becomes WC
stars but none becomes WO. At Z=0.004, only stars with M $>$ 60 $M_\odot$
become WC stars. The 60 $M_\odot$ becomes a WO star.
}
\label{wo1}
\end{figure}
\begin{figure}[!tbp]
\centering
\includegraphics[width=8.5cm]{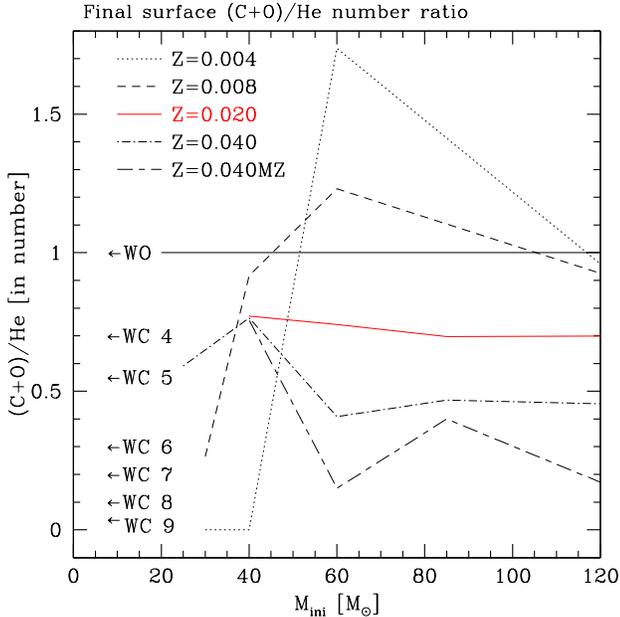}
\caption{Final surface(C+O)/He number ratio as a function of the initial 
mass of the models at different metallicities. WO stars are only found at
the metallicities of the Magellanic clouds.
}
\label{miwo}
\end{figure}
In the present grid of models, no WO star appears at solar or higher 
metallicities (see Table \ref{WCWO} and Fig. \ref{wo1}).
In Fig. \ref{miwo}, we can roughly estimate the mass ranges of WO star
formation at the different metallicities. The rate of WO stars and the
mass ranges are given in Table \ref{grbrates}. Combining all the criteria
for GRB progenitors, we obtain the predicted rates for GRB production in a
normal galaxy coming from WO stars, 
${\cal R}_{\rm GRB}$(WO) given in Table \ref{grbrates}.

These rates are upper limits because we did not convolve our rates with
the initial rotational velocity distribution. Indeed slow rotators do not
produce GRBs. Very fast rotators may not produce GRBs. As can be
seen in Table \ref{moma04}, the final angular momentum contained in the
model with $Z=Z_{\rm SMC}$, an initial mass of 60 $M_\odot$ and an initial 
rotational velocity of 500\,km\,s$^{-1}$ is not sufficient.
Magnetic braking may also reduce the mass range for GRB production.
However, since the WO stars spend little or no time during the RSG phase
and have shorter lifetimes than stars around 15 $M_\odot$ stars, the
magnetic braking may not be as strong for WO as expected by recent 
calculations by \citet{HWS05}.
Finally, these rates have to be convolved in metallicity to compare with
observations.
The ${\cal R}_{\rm GRB}$(WO) rates are anyway in much better agreement with 
observation 
than the rates from all WR stars.
The very interesting point about WO stars as GRB progenitors is their
metallicity dependence. Indeed there is no WO predicted at solar or higher
metallicities. At very low metallicities, the existence of WO stars
depends on the initial rotational velocity of the stars. If extremely low
metallicity stars have the same initial surface velocity as solar metallicity
stars, no WO stars can form at extremely low metallicity. However, if the
same amount of angular momentum as solar metallicity
stars is contained in extremely low metallicity stars, WO stars may form
(see Meynet, Ekstr\"om and Maeder, in prep.).
Observations of host galaxies of GRB show that these galaxies are
sub--luminous and blue, pointing in the direction that GRB production is
favoured at low metallicity \citep{Fl03} as is predicted in this study.

\section{Conclusion}
We present the evolution of rotation in models of massive rotating single
stars covering
a wide range in mass and metallicity.
These models reproduce very well observations during the early stages of
the evolution \citep[especially WR populations and ratio between type II and type Ib,c at
different metallicities, see][]{ROTXI}.  

Our models predict the production of fast rotating black holes. 
Models with large initial masses or high metallicity end with less angular momentum
in their central remnant with respect to the break--up limit for the remnant.
Many WR star models satisfy the three main criteria (black hole formation,
loss of hydrogen--rich envelope and enough angular momentum to form an
accretion disk around the black hole) for long and soft gamma--ray bursts (GRB) 
production via the collapsar
model \citep{W93}. Considering all types of WR stars as GRB progenitors, 
we predict too many GRBs compared to observations. If we consider only 
WO stars \citep[type
Ic supernovae as is the case for SN2003dh/GRB030329][]{MMN03} as
GRB progenitors, the GRB production rates are in much better agreement
with observations. WO stars are produced only at low metallicities in the
present grid of models. Although the numbers estimated in this work are
to be taken with caution, the trend of the effects found is clear.
The prediction for the metallicity dependence can be tested by future
observations. 

Some recent studies \citep{HWS05,PMN04,IRT04} suggest that single stars 
can not retain enough
angular momentum for GRBs production and predict that binary stars are the
progenitors of GRBs. 
However,
\citet{IRT04} use a simple description for angular momentum and
their model without magnetic braking are not consistent with our models
and those of \citet{HLW00}. 
\citet{FH05} studied massive binary star mergers in detail. 
They found that, in
general, the final angular momentum contained in the 
central remnant is
similar in merged stars and in single massive stars.
\citet{PLYH05} find that, including the effects of magnetic fields according
to \citet{Sp02}, neither the single nor the binary star models that they calculated 
retain enough angular momentum to form GRBs. 
Finally, magnetic braking which is implemented mainly to reproduce the
observed pulsar periods (which is a very important observation), may be much
less efficient for WO stars than for NS forming stars because WO stars
spend little or no time in the RSG stage and have shorter lifetimes.

In conclusion, in the present models, the best candidates for GRB 
progenitors are, for single massive stars, 
Wolf--Rayet stars of the type WO, which are found only at low
metallicities (Z$\sim$Z$_{\rm SMC}$) and have masses, 
$M \gtrsim$ 50 $M_\odot$. In order to support this prediction, 
additional observations are needed to confirm
the link between type Ic supernovae and long--soft GRBs.

\bibliographystyle{aa}

\begin{thebibliography}{40}
\expandafter\ifx\csname natexlab\endcsname\relax\def\natexlab#1{#1}\fi

\bibitem[{{ATNF}(2005)}]{PSR}
{ATNF}. 2005, www.atnf.csiro.au/research/pulsar/psrcat

\bibitem[{{Braithwaite} \& {Spruit}(2004)}]{BS04}
{Braithwaite}, J. \& {Spruit}, H.~C. 2004, \nat, 431, 819

\bibitem[{{Cappellaro} {et~al.}(1999){Cappellaro}, {Evans}, \&
  {Turatto}}]{Ca99}
{Cappellaro}, E., {Evans}, R., \& {Turatto}, M. 1999, \aap, 351, 459

\bibitem[{{Fryer}(1999)}]{F99}
{Fryer}, C.~L. 1999, \apj, 522, 413

\bibitem[{{Fryer} \& {Heger}(2005)}]{FH05}
{Fryer}, C.~L. \& {Heger}, A. 2005, \apj, 623, 302

\bibitem[{{Fukuda}(1982)}]{FU82}
{Fukuda}, I. 1982, \pasp, 94, 271

\bibitem[{{Heger} {et~al.}(2003){Heger}, {Fryer}, {Woosley}, {Langer}, \&
  {Hartmann}}]{HFWLH03}
{Heger}, A., {Fryer}, C.~L., {Woosley}, S.~E., {Langer}, N., \& {Hartmann},
  D.~H. 2003, \apj, 591, 288

\bibitem[{{Heger} {et~al.}(2000){Heger}, {Langer}, \& {Woosley}}]{HLW00}
{Heger}, A., {Langer}, N., \& {Woosley}, S.~E. 2000, \apj, 528, 368

\bibitem[{{Heger} {et~al.}(2004){Heger}, {Woosley}, {Langer}, \&
  {Spruit}}]{HWLS03}
{Heger}, A., {Woosley}, S.~E., {Langer}, N., \& {Spruit}, H.~C. 2004, in IAU
  Symposium 215: Stellar rotation, 591/ astro--ph0301374

\bibitem[{{Heger} {et~al.}(2005){Heger}, {Woosley}, \& {Spruit}}]{HWS05}
{Heger}, A., {Woosley}, S.~E., \& {Spruit}, H.~C. 2005, \apj, 626, 350

\bibitem[{{Hirschi} {et~al.}(2004){Hirschi}, {Meynet}, \& {Maeder}}]{psn04a}
{Hirschi}, R., {Meynet}, G., \& {Maeder}, A. 2004, \aap, 425, 649

\bibitem[{{Izzard} {et~al.}(2004){Izzard}, {Ramirez-Ruiz}, \& {Tout}}]{IRT04}
{Izzard}, R.~G., {Ramirez-Ruiz}, E., \& {Tout}, C.~A. 2004, \mnras, 348, 1215

\bibitem[{{Janka} {et~al.}(2005){Janka}, {Scheck}, {Kifonidis}, {M{\" u}ller},
  \& {Plewa}}]{J05}
{Janka}, H.-T., {Scheck}, L., {Kifonidis}, K., {M{\" u}ller}, E., \& {Plewa},
  T. 2005, in ASP Conference Series, Vol. 332: The Fate of the Most Massive
  Stars, p.372/ astro-ph0408439

\bibitem[{{Kaper} \& {van der Meer}(2005)}]{KM05}
{Kaper}, L. \& {van der Meer}, A. 2005, astro-ph/0502314

\bibitem[{{Kroupa}(2005)}]{Kr05}
{Kroupa}, P. 2005, \nat, 434, 148

\bibitem[{{Lattimer} \& {Prakash}(2001)}]{LP01}
{Lattimer}, J.~M. \& {Prakash}, M. 2001, \apj, 550, 426

\bibitem[{{Le Floc'h} {et~al.}(2003){Le Floc'h}, {Duc}, {Mirabel}, {Sanders},
  {Bosch}, {Diaz}, {Donzelli}, {Rodrigues}, {Courvoisier}, {Greiner},
  {Mereghetti}, {Melnick}, {Maza}, \& {Minniti}}]{Fl03}
{Le Floc'h}, E., {Duc}, P.-A., {Mirabel}, I.~F., {et~al.} 2003, \aap, 400, 499

\bibitem[{{MacFadyen} \& {Woosley}(1999)}]{FW99}
{MacFadyen}, A.~I. \& {Woosley}, S.~E. 1999, \apj, 524, 262

\bibitem[{{Maeder}(1992)}]{AM92}
{Maeder}, A. 1992, \aap, 264, 105

\bibitem[{{Maeder} \& {Meynet}(2000)}]{ROTVI}
{Maeder}, A. \& {Meynet}, G. 2000, \aap, 361, 159

\bibitem[{{Maeder} \& {Meynet}(2001)}]{ROTVII}
---. 2001, \aap, 373, 555

\bibitem[{{Maeder} \& {Meynet}(2004)}]{ROTMII}
---. 2004, \aap, 422, 225

\bibitem[{{Maeder} \& {Meynet}(2005)}]{ROTBIII}
---. 2005, \aap \ in press, astro-ph/0506347

\bibitem[{{Marshall} {et~al.}(1998){Marshall}, {Gotthelf}, {Zhang},
  {Middleditch}, \& {Wang}}]{MG98}
{Marshall}, F.~E., {Gotthelf}, E.~V., {Zhang}, W., {Middleditch}, J., \&
  {Wang}, Q.~D. 1998, \apjl, 499, L179

\bibitem[{{Matheson}(2003)}]{Ma03}
{Matheson}, T. 2003, astro-ph/0309793

\bibitem[{{Mazzali} {et~al.}(2003){Mazzali}, {Deng}, {Tominaga}, {Maeda},
  {Nomoto}, {Matheson}, {Kawabata}, {Stanek}, \& {Garnavich}}]{MMN03}
{Mazzali}, P.~A., {Deng}, J., {Tominaga}, N., {et~al.} 2003, \apjl, 599, L95

\bibitem[{{Meynet} \& {Maeder}(1997)}]{ROTI}
{Meynet}, G. \& {Maeder}, A. 1997, \aap, 321, 465

\bibitem[{{Meynet} \& {Maeder}(2002{\natexlab{a}})}]{ROTVIII}
---. 2002{\natexlab{a}}, \aap, 390, 561

\bibitem[{{Meynet} \& {Maeder}(2002{\natexlab{b}})}]{MM02n}
---. 2002{\natexlab{b}}, \aap, 381, L25

\bibitem[{{Meynet} \& {Maeder}(2003)}]{ROTX}
---. 2003, \aap, 404, 975

\bibitem[{{Meynet} \& {Maeder}(2005)}]{ROTXI}
---. 2005, \aap, 429, 581

\bibitem[{{Morrison} {et~al.}(2004){Morrison}, {Baumgarte}, \&
  {Shapiro}}]{MBS04}
{Morrison}, I.~A., {Baumgarte}, T.~W., \& {Shapiro}, S.~L. 2004, \apj, 610, 941

\bibitem[{{Ott} {et~al.}(2005){Ott}, {Ou}, {Tohline}, \& {Burrows}}]{OTB05}
{Ott}, C.~D., {Ou}, S., {Tohline}, J.~E., \& {Burrows}, A. 2005, \apjl, 625,
  L119

\bibitem[{{Petrovic} {et~al.}(2005){Petrovic}, {Langer}, {Yoon}, \&
  {Heger}}]{PLYH05}
{Petrovic}, J., {Langer}, N., {Yoon}, S.-C., \& {Heger}, A. 2005, \aap, 435,
  247

\bibitem[{{Piran}(2005)}]{Pi04}
{Piran}, T. 2005, Reviews of Modern Physics, 76, 1143

\bibitem[{{Podsiadlowski} {et~al.}(2004){Podsiadlowski}, {Mazzali}, {Nomoto},
  {Lazzati}, \& {Cappellaro}}]{PMN04}
{Podsiadlowski}, P., {Mazzali}, P.~A., {Nomoto}, K., {Lazzati}, D., \&
  {Cappellaro}, E. 2004, \apjl, 607, L17

\bibitem[{{Salpeter}(1955)}]{Sa55}
{Salpeter}, E.~E. 1955, \apj, 121, 161

\bibitem[{{Seward} {et~al.}(1984){Seward}, {Harnden}, \& {Helfand}}]{SHH84}
{Seward}, F.~D., {Harnden}, F.~R., \& {Helfand}, D.~J. 1984, \apjl, 287, L19

\bibitem[{{Shapiro} \& {Teukolsky}(1983)}]{ST83}
{Shapiro}, S.~L. \& {Teukolsky}, S.~A. 1983, {Black-Holes White Dwarfs and
  Neutron Stars} (John Wiley \& Sons)

\bibitem[{{Smith} \& {Maeder}(1991)}]{SM91}
{Smith}, L.~F. \& {Maeder}, A. 1991, \aap, 241, 77

\bibitem[{{Spruit}(2002)}]{Sp02}
{Spruit}, H.~C. 2002, \aap, 381, 923

\bibitem[{{Srinivasan}(2002)}]{Sr02}
{Srinivasan}, G. 2002, \aapr, 11, 67

\bibitem[{{Staelin} \& {Reifenstein}(1968)}]{SR68}
{Staelin}, D.~H. \& {Reifenstein}, E.~C. 1968, Science, 162, 1481

\bibitem[{{Vietri} \& {Stella}(1999)}]{VS99}
{Vietri}, M. \& {Stella}, L. 1999, \apjl, 527, L43

\bibitem[{{Wheeler} {et~al.}(2002){Wheeler}, {Meier}, \& {Wilson}}]{WMW02}
{Wheeler}, J.~C., {Meier}, D.~L., \& {Wilson}, J.~R. 2002, \apj, 568, 807

\bibitem[{{Woosley}(1993)}]{W93}
{Woosley}, S.~E. 1993, \apj, 405, 273

\bibitem[{{Woosley} \& {Heger}(2003)}]{WH03}
{Woosley}, S.~E. \& {Heger}, A. 2003, in IAU Symposium 215: Stellar Rotation,
  601/ astro-ph0301373

\end{thebibliography}

\begin{table*}
\caption{Z=0.004 (SMC): angular and momenta of inertia for rotating stellar 
models at various stages of their evolution.
For each model, we give the initial mass and velocity as well as the remnant
mass estimated from the carbon--oxygen (CO) core mass using the relation from
\citet{AM92}. We used the CO core at the end of the evolution for the
calculation: at the end of core He--burning in general and at the end of
Si--burning for the models from paper XII.
Then, column 1 is the evolutionary stage to which the values correspond,
 column 2 is the total mass at the given stage,
 column 3 is angular momentum contained in the remnant,
${\cal L}_{\rm rem}=\int_0^{M_{\rm rem}}2/3\,\Omega\,r^2\,dm$
in units of $10^{50}\,g\,cm^2\,s^{-1}$,
 column 4 is the moment of inertia of the remnant, 
${\cal I}_{\rm rem}=\int_0^{M_{\rm rem}}r^2\,dm$
in units of $10^{55}\,g\,cm^2$,
 column 5 is the specific angular momentum at the remnant edge, $j_{\rm rem}$,
in units of $10^{16}\,cm^2\,s^{-1}$,
 column 6 is the average specific angular momentum in the remnant, 
$\overline{j}_{\rm rem}={\cal L}_{\rm rem}/M_{\rm rem}$, 
in units of $10^{16}\,cm^2\,s^{-1}$,
 column 7 is the specific angular momentum at the mass coordinate 1.56 $M_\odot$, 
$j_{1.56}$, in units of $10^{16}\,cm^2\,s^{-1}$, 
 column 8 is the angular momentum contained in the inner 1.56 $M_\odot$,
${\cal L}_{1.56}=\int_0^{1.56\,M_\odot}2/3\,\Omega\,r^2\,dm$
in units of $10^{50}\,g\,cm^2\,s^{-1}$,
 column 9 is the specific angular momentum at the mass coordinate 2.5 $M_\odot$, 
$j_{2.5}$, in units of $10^{16}\,cm^2\,s^{-1}$, 
 column 10 is the angular momentum contained in the inner 2.5 $M_\odot$,
${\cal L}_{2.5}=\int_0^{2.5\,M_\odot}2/3\,\Omega\,r^2\,dm$
in units of $10^{50}\,g\,cm^2\,s^{-1}$.
Assuming that a neutron star with a baryonic mass, $M_b= 1.56 M_\odot$, and with a 
 radius, $R=12\,$km, would
form from the models, we calculated the two following quantities:
 (column 11) the ratio of the NS angular velocity to its 
Keplerian angular velocity, $\Omega/\Omega_K$(NS).
 (column 12) the initial rotation period of the neutron star,
${\cal P}_{\rm rot}$ in units of milli--seconds.
} \label{moma04}
\scriptsize
\begin{center}
\begin{tabular}{c r r r r r r r r r r r}
\hline
 \hline \\
stage   &   Mass & ${\cal L}_{\rm rem}$ &  ${\cal I}_{\rm rem}$  
  & $j_{\rm rem}$  & ${\overline j}_{\rm rem}$ 
  & $j_{1.56}$ & ${\cal L}_{1.56}$ 
  &  $j_{2.5}$ &  ${\cal L}_{2.5}$ 
  & ($\Omega/\Omega_{\rm K,NS})$ & $({\cal P}_{\rm rot})$   
  \\ \\
        &        & 10$^{50}$ &  10$^{55}$  
  & 10$^{16}$  & 10$^{16}$ 
  & 10$^{16}$ & 10$^{50}$ 
  &  10$^{16}$ &  10$^{50}$ 
  &  &    
  \\
   &  [$M_\odot$] &  [$\frac{{\rm g\,cm}^2}{\rm s}$] &  [g\,cm$^2$]
   & [$\frac{{\rm cm}^2}{{\rm s}}$] & [$\frac{{\rm cm}^2}{{\rm s}}$] 
   & [$\frac{{\rm cm}^2}{{\rm s}}$] & [$\frac{{\rm g\,cm}^2}{\rm s}$] 
   & [$\frac{{\rm cm}^2}{{\rm s}}$] & [$\frac{{\rm g\,cm}^2}{\rm s}$] 
   &   & [ms] \\ \\
 \hline
\multispan{12}{\hfill  30 $M_\odot$, $\upsilon_{\rm ini}$ = 300 km\,s$^{-1}$,$M_{\rm rem}$ = 3.73 M$_\odot$ \hfill} \\
     ZAMS   &    30.0  &  11.970   &   1.8E+00  &  27.13  &  16.14  &  14.24  & 2.875  &  20.07  & 6.190  &  17.20  &   0.035   \\
    end H   &    28.7  &   4.309   &   4.6E-01  &  10.06  &   5.81  &   5.03  & 1.003  &   7.24  & 2.190  &   6.00  &   0.101   \\
 start He   &    28.5  &   4.123   &   9.3E-02  &   9.88  &   5.56  &   4.73  & 0.934  &   6.94  & 2.063  &   5.59  &   0.108   \\
   end He   &    18.9  &   3.153   &   2.8E-02  &   7.50  &   4.25  &   3.64  & 0.683  &   5.30  & 1.534  &   4.09  &   0.148   \\
\multispan{12}{\hfill  40 $M_\odot$, $\upsilon_{\rm ini}$ = 300 km\,s$^{-1}$,$M_{\rm rem}$ = 5.40 M$_\odot$ \hfill} \\
     ZAMS   &    40.0  &  21.882   &   3.8E+00  &  34.63  &  20.39  &  13.86  & 2.789  &  19.45  & 6.164  &  16.69  &   0.036   \\
    end H   &    36.8  &   6.805   &   1.3E+00  &  11.05  &   6.34  &   4.20  & 0.839  &   5.97  & 1.870  &   5.02  &   0.121   \\
 start He   &    36.5  &   6.367   &   2.0E-01  &  10.59  &   5.93  &   3.84  & 0.760  &   5.52  & 1.709  &   4.55  &   0.133   \\
 start WR   &    27.0  &   6.012   &   1.7E-01  &   9.97  &   5.60  &   3.64  & 0.721  &   5.22  & 1.552  &   4.32  &   0.140   \\
   end He   &    22.3  &   5.176   &   5.4E-02  &   8.58  &   4.82  &   3.13  & 0.592  &   4.49  & 1.337  &   3.54  &   0.171   \\
\multispan{12}{\hfill  60 $M_\odot$, $\upsilon_{\rm ini}$ = 300 km\,s$^{-1}$,$M_{\rm rem}$ = 8.52 M$_\odot$ \hfill} \\
     ZAMS   &    60.0  &  44.629   &   9.7E+00  &  44.97  &  26.35  &  12.95  & 2.743  &  18.08  & 5.775  &  16.41  &   0.037   \\
    end H   &    51.8  &  11.797   &   3.2E+00  &  12.16  &   6.97  &   3.32  & 0.700  &   4.67  & 1.482  &   4.19  &   0.145   \\
 start He   &    50.9  &  10.968   &   5.4E-01  &  11.52  &   6.48  &   3.01  & 0.632  &   4.26  & 1.344  &   3.78  &   0.160   \\
 start WR   &    44.2  &  10.697   &   4.9E-01  &  11.23  &   6.32  &   2.94  & 0.617  &   4.16  & 1.312  &   3.69  &   0.164   \\
   end He   &    28.5  &   8.076   &   6.2E-02  &   8.70  &   4.77  &   2.21  & 0.431  &   3.14  & 0.945  &   2.58  &   0.235   \\
\multispan{12}{\hfill  60 $M_\odot$, $\upsilon_{\rm ini}$ = 500 km\,s$^{-1}$,$M_{\rm rem}$ = 3.79 M$_\odot$ \hfill} \\
     ZAMS   &    60.0  &  18.372   &   2.5E+00  &  38.32  &  24.36  &  20.33  & 4.306  &  28.38  & 9.066  &  25.76  &   0.024   \\
 start WR   &    50.3  &   4.734   &   2.1E+00  &   9.96  &   6.28  &   5.21  & 1.098  &   7.32  & 2.322  &   6.57  &   0.092   \\
    end H   &    36.7  &   1.920   &   8.0E-01  &   4.23  &   2.55  &   2.19  & 0.461  &   3.09  & 0.978  &   2.76  &   0.219   \\
 start He   &    36.3  &   1.754   &   1.3E-01  &   3.90  &   2.33  &   1.99  & 0.418  &   2.83  & 0.890  &   2.50  &   0.242   \\
   end He   &    12.4  &   1.102   &   2.8E-02  &   2.63  &   1.46  &   1.24  & 0.228  &   1.82  & 0.518  &   1.37  &   0.443   \\
\multispan{12}{\hfill 120 $M_\odot$, $\upsilon_{\rm ini}$ = 300 km\,s$^{-1}$,$M_{\rm rem}$ = 5.14 M$_\odot$ \hfill} \\
     ZAMS   &   119.9  &  16.778   &   5.4E+00  &  25.59  &  16.40  &  11.09  & 2.580  &  15.37  & 5.408  &  15.44  &   0.039   \\
 start WR   &    97.7  &   5.849   &   4.6E+00  &   8.96  &   5.72  &   3.85  & 0.893  &   5.35  & 1.876  &   5.34  &   0.113   \\
    end H   &    53.0  &   1.065   &   1.3E+00  &   1.73  &   1.04  &   0.72  & 0.154  &   1.01  & 0.323  &   0.92  &   0.655   \\
 start He   &    52.5  &   1.052   &   2.5E-01  &   1.71  &   1.03  &   0.71  & 0.153  &   1.00  & 0.319  &   0.91  &   0.664   \\
   end He   &    17.2  &   0.478   &   5.3E-02  &   0.84  &   0.47  &   0.31  & 0.059  &   0.45  & 0.131  &   0.35  &   1.724   \\
 \hline \\                                                                                                                                                                         
\end{tabular}
\end{center}
\end{table*}

\begin{table*}
\caption{Z=0.008 (LMC): angular and momenta of inertia for rotating stellar models at various
stages of their evolution (see Table \ref{moma04} for the legend).} \label{moma08}
\scriptsize
\begin{center}
\begin{tabular}{c r r r r r r r r r r r}
\hline
 \hline \\
stage   &   Mass & ${\cal L}_{\rm rem}$ &  ${\cal I}_{\rm rem}$  
  & $j_{\rm rem}$  & ${\overline j}_{\rm rem}$ 
  & $j_{1.56}$ & ${\cal L}_{1.56}$ 
  &  $j_{2.5}$ &  ${\cal L}_{2.5}$ 
  & ($\Omega/\Omega_{\rm K,NS})$ & $({\cal P}_{\rm rot})$   
  \\ \\
        &        & 10$^{50}$ &  10$^{55}$  
  & 10$^{16}$  & 10$^{16}$ 
  & 10$^{16}$ & 10$^{50}$ 
  &  10$^{16}$ &  10$^{50}$ 
  &  &    
  \\
   &  [$M_\odot$] &  [$\frac{{\rm g\,cm}^2}{\rm s}$] &  [g\,cm$^2$]
   & [$\frac{{\rm cm}^2}{{\rm s}}$] & [$\frac{{\rm cm}^2}{{\rm s}}$] 
   & [$\frac{{\rm cm}^2}{{\rm s}}$] & [$\frac{{\rm g\,cm}^2}{\rm s}$] 
   & [$\frac{{\rm cm}^2}{{\rm s}}$] & [$\frac{{\rm g\,cm}^2}{\rm s}$] 
   &   & [ms] \\ \\
 \hline
\multispan{12}{\hfill  30 $M_\odot$, $\upsilon_{\rm ini}$ = 300 km\,s$^{-1}$,$M_{\rm rem}$ = 3.73 M$_\odot$ \hfill} \\
     ZAMS   &    30.0  &  12.106   &   1.9E+00  &  27.46  &  16.34  &  14.42  & 2.911  &  20.33  & 6.268  &  17.42  &   0.035   \\
    end H   &    27.5  &   4.072   &   5.3E-01  &   9.51  &   5.49  &   4.76  & 0.949  &   6.85  & 2.073  &   5.68  &   0.107   \\
 start He   &    27.3  &   3.880   &   9.7E-02  &   9.30  &   5.24  &   4.46  & 0.880  &   6.54  & 1.944  &   5.27  &   0.115   \\
 start WR   &    16.3  &   3.365   &   7.9E-02  &   8.13  &   4.54  &   3.96  & 0.744  &   5.77  & 1.670  &   4.45  &   0.136   \\
   end He   &    12.0  &   2.804   &   1.1E-02  &   6.86  &   3.78  &   3.27  & 0.597  &   4.81  & 1.381  &   3.57  &   0.170   \\
\multispan{12}{\hfill  40 $M_\odot$, $\upsilon_{\rm ini}$ = 300 km\,s$^{-1}$,$M_{\rm rem}$ = 5.21 M$_\odot$ \hfill} \\
     ZAMS   &    40.0  &  20.719   &   3.9E+00  &  33.88  &  19.99  &  13.93  & 2.803  &  19.54  & 6.193  &  16.77  &   0.036   \\
    end H   &    35.3  &   5.935   &   1.2E+00  &   9.95  &   5.73  &   3.90  & 0.777  &   5.54  & 1.733  &   4.65  &   0.130   \\
 start He   &    34.8  &   5.510   &   2.0E-01  &   9.45  &   5.32  &   3.54  & 0.701  &   5.08  & 1.575  &   4.19  &   0.144   \\
 start WR   &    28.1  &   5.338   &   1.7E-01  &   9.14  &   5.15  &   3.43  & 0.680  &   4.93  & 1.464  &   4.07  &   0.149   \\
   end He   &    17.2  &   4.052   &   6.5E-03  &   7.11  &   3.91  &   2.62  & 0.493  &   3.78  & 1.118  &   2.95  &   0.205   \\
\multispan{12}{\hfill  60 $M_\odot$, $\upsilon_{\rm ini}$ = 300 km\,s$^{-1}$,$M_{\rm rem}$ = 4.91 M$_\odot$ \hfill} \\
     ZAMS   &    60.0  &  17.965   &   4.1E+00  &  29.67  &  18.40  &  13.02  & 2.757  &  18.17  & 5.804  &  16.49  &   0.037   \\
    end H   &    48.4  &   5.156   &   7.7E-01  &   8.64  &   5.28  &   3.68  & 0.775  &   5.18  & 1.642  &   4.64  &   0.131   \\
 start He   &    46.9  &   4.594   &   2.1E-01  &   8.02  &   4.71  &   3.36  & 0.705  &   4.75  & 1.499  &   4.22  &   0.143   \\
 start WR   &    43.8  &   4.523   &   1.8E-01  &   7.90  &   4.63  &   3.31  & 0.694  &   4.68  & 1.476  &   4.16  &   0.146   \\
   end He   &    16.4  &   2.688   &   4.1E-02  &   4.98  &   2.75  &   1.92  & 0.358  &   2.78  & 0.804  &   2.14  &   0.283   \\
\multispan{12}{\hfill 120 $M_\odot$, $\upsilon_{\rm ini}$ = 300 km\,s$^{-1}$,$M_{\rm rem}$ = 4.06 M$_\odot$ \hfill} \\
     ZAMS   &   119.9  &  11.385   &   4.0E+00  &  21.45  &  14.11  &  11.00  & 2.560  &  15.25  & 5.366  &  15.32  &   0.040   \\
 start WR   &    93.8  &   3.394   &   3.5E+00  &   6.42  &   4.21  &   3.27  & 0.759  &   4.55  & 1.595  &   4.54  &   0.133   \\
    end H   &    36.6  &   0.525   &   7.5E-01  &   1.07  &   0.65  &   0.53  & 0.106  &   0.74  & 0.235  &   0.63  &   0.958   \\
 start He   &    36.3  &   0.519   &   1.6E-01  &   1.06  &   0.64  &   0.52  & 0.104  &   0.73  & 0.232  &   0.62  &   0.970   \\
   end He   &    13.3  &   0.277   &   6.3E-03  &   0.62  &   0.34  &   0.28  & 0.052  &   0.41  & 0.116  &   0.31  &   1.950   \\
 \hline \\
\end{tabular}
\end{center}
\end{table*}

\begin{table*}
\caption{Z=0.020 (solar): angular and momenta of inertia for rotating stellar models at various
stages of their evolution (see Table \ref{moma04} for the legend).} \label{moma20}
\scriptsize
\begin{center}
\begin{tabular}{c r r r r r r r r r r r}
\hline
 \hline \\
stage   &   Mass & ${\cal L}_{\rm rem}$ &  ${\cal I}_{\rm rem}$  
  & $j_{\rm rem}$  & ${\overline j}_{\rm rem}$ 
  & $j_{1.56}$ & ${\cal L}_{1.56}$ 
  &  $j_{2.5}$ &  ${\cal L}_{2.5}$ 
  & ($\Omega/\Omega_{\rm K,NS})$ & $({\cal P}_{\rm rot})$   
  \\ \\
        &        & 10$^{50}$ &  10$^{55}$  
  & 10$^{16}$  & 10$^{16}$ 
  & 10$^{16}$ & 10$^{50}$ 
  &  10$^{16}$ &  10$^{50}$ 
  &  &    
  \\
   &  [$M_\odot$] &  [$\frac{{\rm g\,cm}^2}{\rm s}$] &  [g\,cm$^2$]
   & [$\frac{{\rm cm}^2}{{\rm s}}$] & [$\frac{{\rm cm}^2}{{\rm s}}$] 
   & [$\frac{{\rm cm}^2}{{\rm s}}$] & [$\frac{{\rm g\,cm}^2}{\rm s}$] 
   & [$\frac{{\rm cm}^2}{{\rm s}}$] & [$\frac{{\rm g\,cm}^2}{\rm s}$] 
   &   & [ms] \\ \\
 \hline
\multispan{12}{\hfill   9 $M_\odot$, $\upsilon_{\rm ini}$ = 300 km\,s$^{-1}$,$M_{\rm rem}$ = 1.18 M$_\odot$ \hfill} \\
     ZAMS   &     9.0  &   1.632   &   1.5E-01  &  11.51  &   6.95  &  14.30  & 2.592  &  21.10  & 5.924  &  15.51  &   0.039   \\
    end H   &     8.8  &   0.744   &   7.1E-02  &   5.49  &   3.17  &   7.07  & 1.196  &  11.90  & 2.976  &   7.16  &   0.085   \\
 start He   &     8.8  &   0.719   &   4.5E-03  &   5.38  &   3.06  &   7.47  & 1.189  &   0.71  & 2.806  &   7.12  &   0.085   \\
   end He   &     8.4  &   0.639   &   2.1E-03  &   5.14  &   2.72  &   5.75  & 1.057  &  13.14  & 2.640  &   6.32  &   0.096   \\
\multispan{12}{\hfill  12 $M_\odot$, $\upsilon_{\rm ini}$ = 300 km\,s$^{-1}$,$M_{\rm rem}$ = 1.46 M$_\odot$ \hfill} \\
     ZAMS   &    12.0  &   2.446   &   2.7E-01  &  13.86  &   8.42  &  14.57  & 2.730  &  21.15  & 6.063  &  16.33  &   0.037   \\
    end H   &    11.5  &   1.025   &   1.1E-01  &   6.05  &   3.53  &   6.40  & 1.149  &   9.92  & 2.642  &   6.88  &   0.088   \\
 start He   &    11.5  &   1.021   &   9.4E-03  &   5.98  &   3.51  &   6.33  & 1.144  &  10.23  & 2.635  &   6.85  &   0.088   \\
   end He   &    10.3  &   0.863   &   3.8E-03  &   5.46  &   2.97  &   5.84  & 0.975  &   8.33  & 2.325  &   5.83  &   0.104   \\
  start O   &    10.2  &   0.688   &   4.2E-05  &   5.53  &   2.37  &   6.86  & 0.821  &   8.63  & 2.141  &   4.91  &   0.123   \\
\multispan{12}{\hfill  15 $M_\odot$, $\upsilon_{\rm ini}$ = 300 km\,s$^{-1}$,$M_{\rm rem}$ = 1.85 M$_\odot$ \hfill} \\
     ZAMS   &    15.0  &   3.682   &   4.5E-01  &  16.79  &  10.01  &  14.76  & 2.762  &  21.21  & 6.196  &  16.52  &   0.037   \\
    end H   &    14.2  &   1.455   &   2.0E-01  &   6.87  &   3.95  &   5.95  & 1.080  &   9.02  & 2.506  &   6.46  &   0.094   \\
 start He   &    14.2  &   1.452   &   1.8E-02  &   6.87  &   3.95  &   6.19  & 1.073  &   8.96  & 2.500  &   6.42  &   0.094   \\
   end He   &    10.4  &   1.182   &   6.6E-03  &   5.78  &   3.21  &   4.93  & 0.872  &   7.88  & 2.065  &   5.22  &   0.116   \\
   end Si   &    10.3  &   0.941   &   1.9E-05  &   4.47  &   2.56  &   4.10  & 0.729  &  11.40  & 2.015  &   4.36  &   0.139   \\
\multispan{12}{\hfill  20 $M_\odot$, $\upsilon_{\rm ini}$ = 300 km\,s$^{-1}$,$M_{\rm rem}$ = 2.57 M$_\odot$ \hfill} \\
     ZAMS   &    20.0  &   6.590   &   9.3E-01  &  21.73  &  12.91  &  14.95  & 2.896  &  21.30  & 6.304  &  17.33  &   0.035   \\
    end H   &    18.2  &   2.322   &   3.5E-01  &   7.92  &   4.55  &   5.25  & 0.997  &   7.74  & 2.217  &   5.96  &   0.102   \\
 start He   &    17.5  &   2.208   &   3.7E-02  &   7.81  &   4.33  &   4.97  & 0.924  &   7.61  & 2.105  &   5.53  &   0.109   \\
   end He   &     8.8  &   1.781   &   1.3E-02  &   6.31  &   3.49  &   4.08  & 0.740  &   6.15  & 1.698  &   4.43  &   0.137   \\
   end Si   &     8.8  &   1.652   &   6.6E-05  &   6.61  &   3.24  &   2.80  & 0.627  &   6.50  & 1.565  &   3.75  &   0.161   \\
\multispan{12}{\hfill  25 $M_\odot$, $\upsilon_{\rm ini}$ = 300 km\,s$^{-1}$,$M_{\rm rem}$ = 3.06 M$_\odot$ \hfill} \\
     ZAMS   &    25.0  &   8.990   &   1.4E+00  &  24.85  &  14.79  &  15.06  & 2.976  &  21.33  & 6.403  &  17.81  &   0.034   \\
    end H   &    21.8  &   2.898   &   5.1E-01  &   8.24  &   4.77  &   4.80  & 0.935  &   6.97  & 2.045  &   5.60  &   0.108   \\
 start He   &    21.6  &   2.786   &   6.5E-02  &   8.19  &   4.58  &   4.55  & 0.873  &   6.80  & 1.944  &   5.22  &   0.116   \\
 start WR   &    13.5  &   2.457   &   5.3E-02  &   7.26  &   4.04  &   4.11  & 0.759  &   6.07  & 1.714  &   4.54  &   0.133   \\
   end He   &    10.2  &   2.143   &   2.1E-02  &   6.38  &   3.53  &   3.57  & 0.658  &   5.31  & 1.492  &   3.94  &   0.154   \\
   end Si   &    10.0  &   1.644   &   7.1E-05  &   3.72  &   2.71  &   1.90  & 0.600  &   3.20  & 1.186  &   3.59  &   0.169   \\
\multispan{12}{\hfill  40 $M_\odot$, $\upsilon_{\rm ini}$ = 300 km\,s$^{-1}$,$M_{\rm rem}$ = 3.85 M$_\odot$ \hfill} \\
     ZAMS   &    40.0  &  12.549   &   2.6E+00  &  26.82  &  16.38  &  13.88  & 2.793  &  19.48  & 6.173  &  16.71  &   0.036   \\
    end H   &    32.9  &   3.035   &   8.9E-01  &   6.61  &   3.96  &   3.32  & 0.662  &   4.71  & 1.476  &   3.96  &   0.153   \\
 start He   &    32.3  &   2.847   &   1.3E-01  &   6.30  &   3.72  &   3.07  & 0.609  &   4.42  & 1.348  &   3.65  &   0.166   \\
 start WR   &    23.9  &   2.659   &   1.0E-01  &   6.02  &   3.47  &   2.95  & 0.586  &   4.24  & 1.260  &   3.51  &   0.173   \\
   end He   &    12.7  &   1.976   &   3.0E-02  &   4.68  &   2.58  &   2.18  & 0.407  &   3.19  & 0.910  &   2.44  &   0.249   \\
   end Si   &    12.6  &   1.399   &   2.9E-04  &   3.00  &   1.83  &   1.86  & 0.361  &   0.98  & 0.849  &   2.16  &   0.281   \\
\multispan{12}{\hfill  60 $M_\odot$, $\upsilon_{\rm ini}$ = 300 km\,s$^{-1}$,$M_{\rm rem}$ = 4.32 M$_\odot$ \hfill} \\
     ZAMS   &    60.0  &  14.451   &   3.8E+00  &  26.77  &  16.81  &  12.90  & 2.732  &  18.01  & 5.753  &  16.35  &   0.037   \\
 start WR   &    44.4  &   3.644   &   3.1E+00  &   7.08  &   4.24  &   3.34  & 0.705  &   4.70  & 1.492  &   4.22  &   0.144   \\
    end H   &    31.4  &   2.548   &   7.2E-01  &   5.01  &   2.96  &   2.32  & 0.454  &   3.28  & 0.991  &   2.71  &   0.223   \\
   end He   &    14.6  &   1.662   &   2.0E-02  &   3.49  &   1.93  &   1.49  & 0.276  &   2.16  & 0.622  &   1.65  &   0.367   \\
   end Si   &    14.6  &   1.193   &   5.0E-04  &   1.71  &   1.39  &   1.00  & 0.250  &   2.17  & 0.615  &   1.50  &   0.404   \\
\multispan{12}{\hfill  85 $M_\odot$, $\upsilon_{\rm ini}$ = 300 km\,s$^{-1}$,$M_{\rm rem}$ = 3.78 M$_\odot$ \hfill} \\
     ZAMS   &    84.9  &  10.546   &   3.4E+00  &  21.97  &  14.04  &  11.79  & 2.638  &  16.39  & 5.461  &  15.78  &   0.038   \\
 start WR   &    64.5  &   4.401   &   3.1E+00  &   9.22  &   5.86  &   4.90  & 1.094  &   6.84  & 2.271  &   6.54  &   0.093   \\
    end H   &    23.6  &   0.727   &   5.6E-01  &   1.67  &   0.97  &   0.85  & 0.161  &   1.21  & 0.357  &   0.97  &   0.627   \\
   end He   &    12.3  &   0.528   &   1.4E-02  &   1.27  &   0.70  &   0.60  & 0.110  &   0.88  & 0.253  &   0.66  &   0.917   \\
\multispan{12}{\hfill 120 $M_\odot$, $\upsilon_{\rm ini}$ = 300 km\,s$^{-1}$,$M_{\rm rem}$ = 3.54 M$_\odot$ \hfill} \\
     ZAMS   &   119.9  &   8.802   &   3.5E+00  &  18.68  &  12.50  &  10.55  & 2.455  &  14.63  & 5.146  &  14.69  &   0.041   \\
 start WR   &    87.2  &   2.344   &   3.2E+00  &   4.99  &   3.33  &   2.80  & 0.650  &   3.90  & 1.367  &   3.89  &   0.156   \\
    end H   &    26.2  &   0.386   &   6.1E-01  &   0.93  &   0.55  &   0.50  & 0.099  &   0.71  & 0.214  &   0.59  &   1.026   \\
 start He   &    26.0  &   0.376   &   1.2E-01  &   0.91  &   0.53  &   0.49  & 0.096  &   0.69  & 0.208  &   0.57  &   1.055   \\
   end He   &    11.3  &   0.248   &   1.6E-02  &   0.64  &   0.35  &   0.32  & 0.059  &   0.47  & 0.133  &   0.35  &   1.719   \\
 \hline \\                        
\end{tabular}
\end{center}
\end{table*}

\begin{table*}
\caption{Z=0.040 (GC): angular and momenta of inertia for rotating stellar models at various
stages of their evolution (see Table \ref{moma04} for the legend).} \label{moma40}
\scriptsize
\begin{center}
\begin{tabular}{c r r r r r r r r r r r}
\hline
 \hline \\
stage   &   Mass & ${\cal L}_{\rm rem}$ &  ${\cal I}_{\rm rem}$  
  & $j_{\rm rem}$  & ${\overline j}_{\rm rem}$ 
  & $j_{1.56}$ & ${\cal L}_{1.56}$ 
  &  $j_{2.5}$ &  ${\cal L}_{2.5}$ 
  & ($\Omega/\Omega_{\rm K,NS})$ & $({\cal P}_{\rm rot})$   
  \\ \\
        &        & 10$^{50}$ &  10$^{55}$  
  & 10$^{16}$  & 10$^{16}$ 
  & 10$^{16}$ & 10$^{50}$ 
  &  10$^{16}$ &  10$^{50}$ 
  &  &    
  \\
   &  [$M_\odot$] &  [$\frac{{\rm g\,cm}^2}{\rm s}$] &  [g\,cm$^2$]
   & [$\frac{{\rm cm}^2}{{\rm s}}$] & [$\frac{{\rm cm}^2}{{\rm s}}$] 
   & [$\frac{{\rm cm}^2}{{\rm s}}$] & [$\frac{{\rm g\,cm}^2}{\rm s}$] 
   & [$\frac{{\rm cm}^2}{{\rm s}}$] & [$\frac{{\rm g\,cm}^2}{\rm s}$] 
   &   & [ms] \\ \\
 \hline
\multispan{12}{\hfill  20 $M_\odot$, $\upsilon_{\rm ini}$ = 300 km\,s$^{-1}$,$M_{\rm rem}$ = 2.68 M$_\odot$ \hfill} \\
     ZAMS   &    20.0  &   7.033   &   1.1E+00  &  22.27  &  13.19   & 14.82  & 2.869  &  21.11  & 6.245  &  17.17  &   0.035   \\
    end H   &    16.9  &   2.223   &   3.0E-01  &   7.29  &   4.17   &  4.66  & 0.882  &   6.87  & 1.966  &   5.28  &   0.115   \\
 start He   &    16.8  &   2.176   &   4.8E-02  &   7.43  &   4.08   &  4.52  & 0.840  &   6.93  & 1.915  &   5.03  &   0.120   \\
   end He   &     9.2  &   1.742   &   1.1E-02  &   5.92  &   3.27   &  3.69  & 0.670  &   5.55  & 1.535  &   4.01  &   0.151   \\
\multispan{12}{\hfill  25 $M_\odot$, $\upsilon_{\rm ini}$ = 300 km\,s$^{-1}$,$M_{\rm rem}$ = 3.13 M$_\odot$ \hfill} \\
     ZAMS   &    25.0  &   9.300   &   1.6E+00  &  25.26  &  14.96   & 15.04  & 2.977  &  21.31  & 6.396  &  17.81  &   0.034   \\
    end H   &    21.3  &   1.980   &   3.2E-01  &   5.48  &   3.18   &  3.18  & 0.623  &   4.58  & 1.352  &   3.73  &   0.163   \\
 start He   &    20.9  &   1.872   &   8.4E-02  &   5.24  &   3.01   &  2.99  & 0.583  &   4.35  & 1.273  &   3.49  &   0.174   \\
 start WR   &    18.5  &   1.854   &   6.6E-02  &   5.18  &   2.98   &  2.96  & 0.554  &   4.30  & 1.249  &   3.31  &   0.183   \\
   end He   &     9.6  &   1.471   &   1.9E-02  &   4.31  &   2.37   &  2.35  & 0.424  &   3.51  & 0.976  &   2.54  &   0.239   \\
\multispan{12}{\hfill  40 $M_\odot$, $\upsilon_{\rm ini}$ = 300 km\,s$^{-1}$,$M_{\rm rem}$ = 3.57 M$_\odot$ \hfill} \\
     ZAMS   &    40.0  &  10.867   &   2.5E+00  &  24.86  &  15.30   & 13.61  & 2.737  &  19.10  & 6.051  &  16.38  &   0.037   \\
 start WR   &    31.1  &   2.400   &   1.7E+00  &   5.74  &   3.38   &  3.07  & 0.612  &   4.35  & 1.308  &   3.66  &   0.165   \\
    end H   &    29.9  &   2.343   &   5.5E-01  &   5.62  &   3.30   &  2.99  & 0.596  &   4.25  & 1.275  &   3.57  &   0.170   \\
 start He   &    29.7  &   2.170   &   1.2E-01  &   5.27  &   3.05   &  2.75  & 0.545  &   3.95  & 1.172  &   3.26  &   0.186   \\
   end He   &    11.4  &   1.516   &   2.3E-02  &   3.87  &   2.13   &  1.91  & 0.356  &   2.81  & 0.798  &   2.13  &   0.285   \\
\multispan{12}{\hfill  40 $M_\odot$, $\upsilon_{\rm ini}$ = 300 km\,s$^{-1}$,$M_{\rm rem}$ = 3.02 M$_\odot$, $\dot{M}_{\rm WR}(Z)$ \hfill} \\
     ZAMS   &    40.0  &   8.250   &   1.9E+00  &  21.95  &  13.72   & 13.61  & 2.737  &  19.10  & 6.051  &  16.38  &   0.037   \\
 start WR   &    31.1  &   1.804   &   1.3E+00  &   5.03  &   3.00   &  3.07  & 0.612  &   4.35  & 1.308  &   3.66  &   0.165   \\
    end H   &    29.5  &   1.754   &   4.0E-01  &   4.91  &   2.92   &  2.98  & 0.594  &   4.24  & 1.271  &   3.55  &   0.170   \\
 start He   &    29.2  &   1.623   &   8.9E-02  &   4.59  &   2.70   &  2.75  & 0.544  &   3.94  & 1.172  &   3.26  &   0.186   \\
   end He   &     9.0  &   1.087   &   1.7E-02  &   3.29  &   1.81   &  1.84  & 0.331  &   2.76  & 0.770  &   1.98  &   0.305   \\
\multispan{12}{\hfill  60 $M_\odot$, $\upsilon_{\rm ini}$ = 300 km\,s$^{-1}$,$M_{\rm rem}$ = 1.94 M$_\odot$, $\dot{M}_{\rm WR}(Z)$ \hfill} \\
 start WR   &    44.9  &   0.971   &   9.9E-01  &   3.83  &   2.52   &  3.27  & 0.691  &   4.60  & 1.461  &   4.13  &   0.146   \\
    end H   &     9.3  &   0.169   &   1.3E-01  &   0.76  &   0.44   &  0.63  & 0.116  &   0.94  & 0.265  &   0.69  &   0.872   \\
 start He   &     9.1  &   0.166   &   3.0E-02  &   0.75  &   0.43   &  0.62  & 0.114  &   0.93  & 0.261  &   0.68  &   0.889   \\
   end He   &     4.8  &   0.136   &   5.5E-03  &   0.66  &   0.35   &  0.53  & 0.091  &   0.87  & 0.221  &   0.54  &   1.111   \\
\multispan{12}{\hfill  85 $M_\odot$, $\upsilon_{\rm ini}$ = 300 km\,s$^{-1}$,$M_{\rm rem}$ = 2.57 M$_\odot$ \hfill} \\
     ZAMS   &    84.8  &   5.316   &   1.9E+00  &  15.57  &  10.40   & 10.98  & 2.456  &  15.27  & 5.086  &  14.70  &   0.041   \\
 start WR   &    62.6  &   2.056   &   1.8E+00  &   6.05  &   4.02   &  4.24  & 0.947  &   5.93  & 1.967  &   5.67  &   0.107   \\
    end H   &    15.6  &   0.334   &   2.5E-01  &   1.12  &   0.65   &  0.75  & 0.142  &   1.09  & 0.318  &   0.85  &   0.711   \\
 start He   &    15.3  &   0.329   &   5.7E-02  &   1.11  &   0.64   &  0.74  & 0.140  &   1.09  & 0.314  &   0.84  &   0.725   \\
   end He   &     7.3  &   0.237   &   1.2E-02  &   0.85  &   0.46   &  0.54  & 0.096  &   0.83  & 0.225  &   0.58  &   1.051   \\
\multispan{12}{\hfill  85 $M_\odot$, $\upsilon_{\rm ini}$ = 300 km\,s$^{-1}$,$M_{\rm rem}$ = 1.95 M$_\odot$, $\dot{M}_{\rm WR}(Z)$ \hfill} \\
     ZAMS   &    84.8  &   3.455   &   1.3E+00  &  12.84  &   8.90   & 10.98  & 2.456  &  15.27  & 5.086  &  14.70  &   0.041   \\
 start WR   &    62.6  &   1.334   &   1.2E+00  &   4.97  &   3.44   &  4.24  & 0.947  &   5.93  & 1.967  &   5.67  &   0.107   \\
    end H   &    11.6  &   0.202   &   1.5E-01  &   0.90  &   0.52   &  0.75  & 0.137  &   1.11  & 0.313  &   0.82  &   0.737   \\
 start He   &    11.3  &   0.204   &   3.0E-02  &   0.92  &   0.53   &  0.76  & 0.138  &   1.14  & 0.317  &   0.82  &   0.735   \\
   end He   &     4.7  &   0.155   &   7.2E-03  &   0.75  &   0.40   &  0.60  & 0.102  &   0.98  & 0.249  &   0.61  &   0.993   \\
\multispan{12}{\hfill 120 $M_\odot$, $\upsilon_{\rm ini}$ = 300 km\,s$^{-1}$,$M_{\rm rem}$ = 2.53 M$_\odot$ \hfill} \\
     ZAMS   &   118.8  &   5.616   &   2.2E+00  &  15.80  &  11.15   & 11.29  & 2.626  &  15.65  & 5.505  &  15.71  &   0.039   \\
 start WR   &    86.5  &   1.314   &   2.1E+00  &   3.71  &   2.61   &  2.64  & 0.613  &   3.67  & 1.287  &   3.66  &   0.165   \\
    end H   &    14.0  &   0.132   &   3.1E-01  &   0.45  &   0.26   &  0.31  & 0.058  &   0.45  & 0.129  &   0.35  &   1.755   \\
 start He   &    13.8  &   0.130   &   5.7E-02  &   0.45  &   0.26   &  0.30  & 0.057  &   0.44  & 0.127  &   0.34  &   1.784   \\
   end He   &     7.1  &   0.103   &   1.3E-02  &   0.38  &   0.21   &  0.24  & 0.043  &   0.37  & 0.101  &   0.26  &   2.341   \\
\multispan{12}{\hfill 120 $M_\odot$, $\upsilon_{\rm ini}$ = 300 km\,s$^{-1}$,$M_{\rm rem}$ = 1.96 M$_\odot$, $\dot{M}_{\rm WR}(Z)$ \hfill} \\
     ZAMS   &   119.6  &   3.969   &   1.5E+00  &  13.68  &  10.16   & 11.68  & 2.717  &  16.20  & 5.697  &  16.26  &   0.037   \\
 start WR   &    86.5  &   0.896   &   1.4E+00  &   3.10  &   2.29   &  2.64  & 0.613  &   3.67  & 1.288  &   3.67  &   0.165   \\
    end H   &     9.6  &   0.083   &   1.6E-01  &   0.37  &   0.21   &  0.30  & 0.055  &   0.45  & 0.127  &   0.33  &   1.842   \\
 start He   &     9.4  &   0.082   &   3.1E-02  &   0.36  &   0.21   &  0.30  & 0.054  &   0.45  & 0.126  &   0.32  &   1.875   \\
   end He   &     4.8  &   0.068   &   6.8E-03  &   0.32  &   0.17   &  0.26  & 0.044  &   0.42  & 0.108  &   0.26  &   2.296   \\
 \hline \\
\end{tabular}
\end{center}
\end{table*}

\end{document}